\documentclass[aps,prb,onecolumn]{revtex4-2}

\usepackage{amsmath,bbm,amsfonts,amsthm,graphicx,color,braket,qcircuit,tikz,multirow}
\usepackage[colorlinks=true,breaklinks=true,linkcolor=magenta,citecolor=magenta,urlcolor=magenta]{hyperref}
\usepackage[version=4]{mhchem}

\newcommand{\hadamard}{\mathrm{Had}}
\newcommand{\CNOT}{\mathsf{CNOT}}
\newcommand*\vett[1]{{\bf{#1}}}
\newcommand*\crt[1]{\hat{a}^\dagger_{#1}}
\newcommand*\dst[1]{\hat{a}^{\phantom{\dagger}}_{#1}}

\newcommand{\important}[1]{{\textbf{#1}}}

\begin{document}

\title{Emerging quantum computing algorithms for quantum chemistry}

\author{Mario Motta}
\thanks{corresponding author, e-mail: mario.motta@ibm.com}
\affiliation{IBM Quantum, IBM Research-Almaden, San Jose, CA 95120, USA}

\author{Julia E. Rice}
\thanks{corresponding author, e-mail: jrice@us.ibm.com}
\affiliation{IBM Quantum, IBM Research-Almaden, San Jose, CA 95120, USA}

\begin{abstract}
Digital quantum computers provide a computational framework for solving the Schr\"{o}dinger equation {for} a variety of many-particle systems.
Quantum computing algorithms for the quantum simulation of these
systems have recently witnessed remarkable growth, {notwithstanding the limitations of existing quantum hardware},  especially as a tool for electronic structure computations in molecules.
In this review, we provide a self-contained introduction to {emerging} algorithms for the simulation of Hamiltonian dynamics and eigenstates,
with emphasis on their applications to the electronic structure in molecular systems.
Theoretical foundations and implementation details of the method are discussed, and their strengths, limitations, and recent advances
are presented.
\end{abstract}

{Article Type: ADVANCED REVIEW}

\maketitle
{\vspace{-1cm}
\small 
\tableofcontents}

\section{Introduction}

Determining the quantum-mechanical behavior of many interacting particles, by means of accurate and predictive computations, is a problem of conceptual and technological relevance \cite{dirac1928quantum}.
An important area within the quantum-mechanical many-body problem 
is represented by molecular chemistry which, over the last few
decades, has been addressed using numerical methods, and designed 
and implemented for a variety of computational platforms
\cite{bartlett2007coupled,helgaker2012recent,helgaker2014molecular}.

More recently, digital quantum computers have been proposed as an alternative and complementary approach to the numerical computation of molecular properties \cite{feynman1982simulating,lloyd1996universal,abrams1997simulation}.
{
Molecular chemistry has been identified as an application for a digital quantum computer. This is because a digital quantum computer can serve as a quantum simulator \cite{georgescu2014quantum} of a molecule, i.e. a controllable quantum system that can be used to study certain properties of a molecule. We will focus on Hamiltonian dynamics, in particular, since at the current state of knowledge, it can be simulated with lower scaling on a digital quantum computer than on a conventional computer  \cite{feynman1982simulating,lloyd1996universal}.}

The idea of a quantum simulator is conceptually interesting and appealing, and the manufacturing, control and operation of quantum-mechanical 
devices is one of the most outstanding open problems in physics. On the other hand, a mutual disconnect exists between the quantum chemistry and quantum information science communities, which represents a significant barrier to progress.
A shared terminology, a rigorous assessment of the potential impact of quantum computers on practical applications, 
including a careful identification of areas where quantum technologies can be relevant, and an appreciation of the subtle complications of quantum chemical research, are necessary for quantum information scientists to conduct research in the quantum simulation of chemistry.
On the other hand, a robust understanding of quantum information science, quantum computational complexity, quantum simulation algorithms, and of the nature and peculiarities of quantum devices
are necessary for chemists to contribute to the design and implementation of algorithms for the quantum simulation of chemistry.

In this work, we aim at bridging the gap between the quantum chemistry and quantum computation communities, by examining 
prospects for quantum computation in molecular chemistry.
We review two important classes of quantum algorithms, 
one for the simulation of Hamiltonian dynamics and the other
for the heuristic simulation of Hamiltonian eigenfunctions.
We analyze their advantages and disadvantages,
suggesting opportunities for future developments 
and synergistic research.

We begin by presenting the concept of simulation and its relevance in quantum chemistry in Section \ref{ref:sec_simulations}. 
We then proceed to describe the concept of a digital quantum computer, providing a high level view to help understand how it can be used {as a simulator} of a quantum system, and focus on strengths and weaknesses of such an approach.

Several important concepts pertaining to the quantum simulation of chemical systems are then presented in Section \ref{sec:algo}.
Emphasis will be placed on the requirements and challenges posed by the adoption of these algorithms in the study of molecular properties.

A discussion about the nature and decoherence phenomena occurring on quantum hardware follows, with the purpose of 
describing error mitigation and correction techniques for today's
quantum hardware. 
Such techniques are a requirement for meaningful calculations on the hardware in the near-term.

Conclusions are drawn in the last section,
highlighting the need of synergistic investigations by quantum information and quantum chemistry scientists 
to understand and overcome the subtle complications of research at the interface between these two fields.
In this context, we will discuss chemical applications that can be targeted to assess and monitor the performance of quantum algorithms and hardware, and opportunities to work collaboratively to improve their performance towards relevant investigations of chemical systems.

Note that in Table \ref{tab:glossary}, the acronyms used throughout this work are listed,
together with their explicit meaning.

\section{Computer simulations in chemistry}
\label{ref:sec_simulations}

\begin{figure}[b!]
\centering
\includegraphics[width=0.88\textwidth]{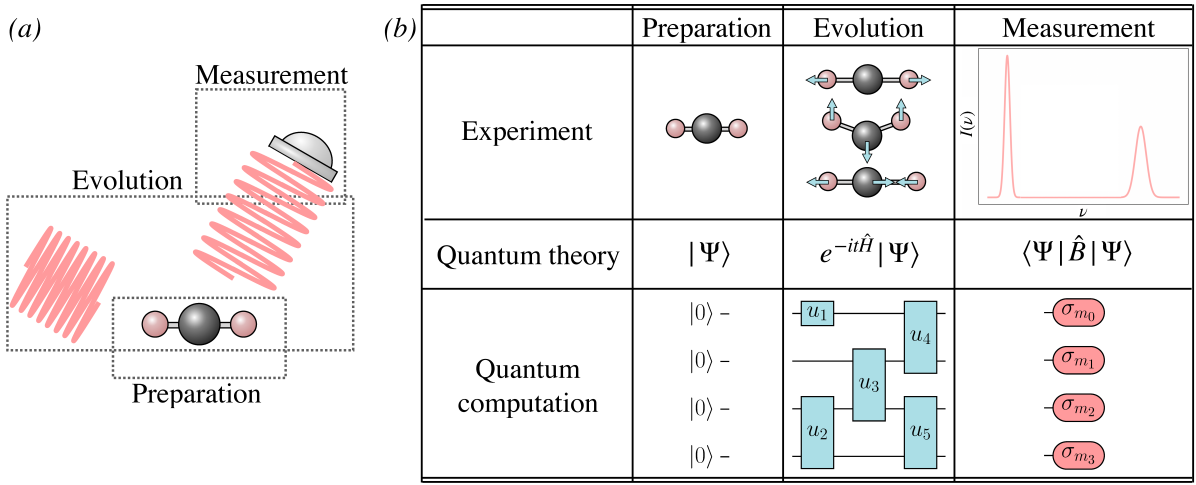}
\caption{a) Schematic representation of infrared spectroscopy experiment on the carbon dioxide molecule. 
b) Analogous phases of both experiment (top row), quantum theory (middle row) and quantum simulations (bottom row).}
\label{fig:experiment}
\end{figure}

Our understanding of the properties of molecular systems often comes from experiments \cite{mukamel1999principles,barron2009molecular}.
A prominent class of experiments is represented by spectroscopic investigations, sketched in Fig.~\ref{fig:experiment}a.
In spectroscopic investigations, electromagnetic radiation is applied to a molecule, and the scattering or absorption of the radiation is measured.
These experimental techniques probe different aspects of the structure of molecules by observing their response to applied electromagnetic fields.
For example, infrared and visible-ultraviolet radiation are used to probe ro-vibrational and electronic excitations, respectively 
\cite{fleming1986chemical,puzzarini2010quantum}; nuclear 
magnetic-resonance spectroscopy to perform in situ identification and concentrations for target chemicals in complex mixtures
\cite{helgaker1999ab};
linear and non-linear optics to probe the polarization of molecules in the presence of external electric fields
\cite{mukamel1999principles}.

Spectroscopic investigations are not the only kinds of experiments performed to understand molecular systems.
Another important class of experiments is represented by chemical reaction rate measurements, 
in which reactants are prepared under suitable experimental conditions, and indicators of chemical compositions, such as electrical conductivity, 
are measured as a function of time, to reveal the evolution of the amount of reactants in the system \cite{hammett1935reaction}.
Such chemical processes can be very complicated
to model, because obtaining accurate rate constants involves description of 
a delicate balance 
of competing phenomena at the molecular level.
These include description of conformers within $k_B \, T$ of the lowest energy conformer, effects of solvation, temperature and pressure, to name but a few.
Even within the Born-Oppenheimer approximation at zero Kelvin, this 
results in potentially very large numbers of intermediates, 
and reaction paths, as in catalytic and metabolic pathways.

Although experiments differ from each other due to a large number of crucially important technical aspects, a recurring theme can be recognized,
which is schematically depicted in Fig.~\ref{fig:experiment}a and in the first row of Fig.~\ref{fig:experiment}b.
A chemical sample is prepared in a suitable quantum state, often at thermal equilibrium at a finite inverse temperature $\beta$.
It is then coupled with an external probe, such as a classical external field, or beam of impinging quantum particles, or a change in a chemical or physical environment, where it evolves under the action of such an external perturbation, and is subsequently measured.

The structure sketched in Fig.~\ref{fig:experiment} is found not 
only in experiments, but also in the theory of quantum mechanics:
an initial preparation is described by wavefunction or density operators in a Hilbert space, the coupling to a probe by a 
unitary transformation or a more general quantum operation,
and a final measurement by a Hermitian operator or a more general
operator-valued measure.
The same structure informs quantum computations, which can thus 
be used to simulate the properties of a quantum system.
The relationship between the phases of an experiment, the postulates
of quantum mechanics, and the structure of a quantum computation
provides a useful high-level framework to recognize the purpose and limitations of quantum computation.

One of the goals of computer simulations in chemistry is to explain and predict the outcome of experiments conducted in laboratories.
Over the last few decades, molecular electronic structure theory has developed to a stage where computational chemistry practitioners can work alongside experimental collaborators to interpret experimental results and to work as a team to design new molecular systems. 
Notwithstanding this progress, the electronic structure of molecules and materials still presents many mysterious aspects,
and methodological developments towards greater accuracy, predictive power, and access to larger systems are to this day highly
investigated research activities.

In the 1980s, the seminal work of several scientists led to the conception of innovative computational devices, now termed digital {(or universal)} quantum computers
\cite{feynman1982simulating,lloyd1996universal,abrams1997simulation}. 
At this point, it is worth remarking that the term "classical" refers to conventional (not quantum) computers. It by no means implies that "classical" is a past era, and in fact, as discussed in the remainder of this work, the best of computations will likely be based on a hybrid approach, where classical and quantum co-processors are used in synergy.

In the context of chemistry, digital quantum computers are used as digital quantum simulators \cite{georgescu2014quantum}. 
By the term \important{quantum simulator}, we denote a controllable quantum system used to simulate the behavior of another quantum system.
{
A \important{digital (or universal) quantum simulator} is a quantum simulator that can be programmed to 
execute any unitary transformation \cite{georgescu2014quantum,bauer2020quantum}.}
The term "simulator" can be a source of confusion.
While the quantum computing literature calls a classical computer emulating 
the behavior of a quantum system a simulator, in quantum information science 
the term refers to an actual quantum system (e.g. an electric circuit with superconducting elements or an array of ions confined and suspended in free 
space using electromagnetic fields) used to execute an algorithm and simulate
the behavior of another quantum system (e.g. a molecule).
We will be using the latter definition in this article.

\begin{figure}[h!]
\centering
\includegraphics[width=0.65\textwidth]{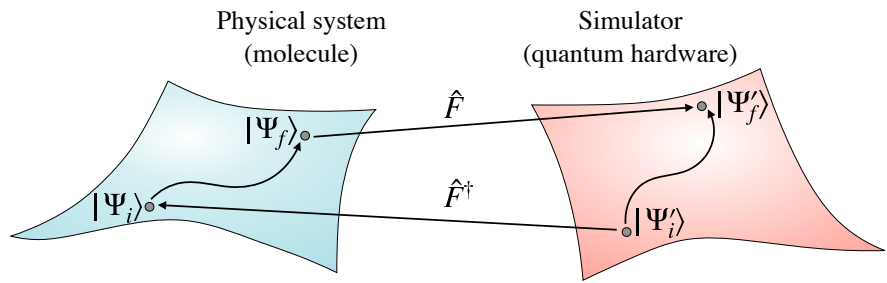}
\caption{Schematic representation of a quantum simulator. {A one-to-one correspondence $\hat{F}$} connects states and operations of a physical system (left, light blue manifold) with states and operations of a controllable quantum system (right, light red manifold) that serves as a simulator for the behavior of the physical system of interest.}
\label{fig:simulation}
\end{figure}

The basic idea of quantum simulation is represented schematically in Fig.~\ref{fig:simulation}. 
We require that wavefunctions in the Hilbert space of the system under study and the simulator can be connected by {a one-to-one correspondence $\hat{F}$, as exemplified in Section \ref{sec:fermions_second}.}
The simulator can then be initialized in a state $| \Psi_i^\prime \rangle$, corresponding to some initial preparation $| \Psi_i \rangle$ of the system under study. 
The simulator can be manipulated with some quantum operation, such as a unitary transformation $\hat{U}^\prime$, 
to reach a final state $| \Psi_f^\prime \rangle$, corresponding  to some state  $| \Psi_f \rangle$ of the system under study. 
The wavefunction $| \Psi_f^\prime \rangle$ has to be measured, to extract information about the state $| \Psi_f \rangle$.

{It is important to emphasize that not all computational problems can be accelerated under a model of computation based on unitary transformations and quantum measurements.}
Thus one must carefully identify areas where the use of quantum simulators can be of actual relevance.
For example, the assumption of controllability of the digital 
quantum simulators is crucial for the actual use of these devices
to solve a computational problem.
This issue will be reviewed in Section \ref{sec:hardware}, where the nature and decoherence phenomena occurring on near-term quantum hardware is discussed, and techniques for error mitigation and correction are presented.
The main factors affecting the performance of quantum hardware
are $(i)$ the limited numbers of qubits that can be used for any one chemical problem, $(ii)$ the limited qubit connectivity, 
and $(iii)$ the various decoherence phenomena that limit the number of quantum operations that can be executed.

Similarly, the importance of preparation, transformation and measurement steps must be stressed here, because the success of a quantum simulation depends 
on the ability to perform certain operations efficiently on the quantum simulator, as well as to extract useful information from it. 
The inherent difficulty of computational problems is typically formalized in terms of the resources required by different  models of computation to solve a given problem.
For digital quantum computers, such a formalization is achieved by the theory of quantum computational complexity, reviewed in Section \ref{sec:complexity},
which identifies the class of problems that are naturally tackled by a digital quantum simulator.
This in turn requires introducing the model of a digital quantum computer and of the operations that can be performed on it, which is the goal of the next Section.

\subsection{Digital quantum computers}
\label{sec:quantum_computers}

The concept of a quantum computer can be represented by several equivalent models, each corresponding to a specific approach to the {problem of executing computation on a device based on quantum mechanics}.
Such models include quantum Turing machines, circuits, random access machines and walks.
In this work, we focus on what is arguably the most widely used model of quantum computation, the circuit model.
It is worth emphasizing that the only educational platforms giving a robust understanding of quantum computation are dedicated textbooks covering the topic in depth and breadth.
Interested readers are thus referred, for example, to the books by Nielsen and Chuang \cite{nielsen2002quantum}, Mermin \cite{mermin2007quantum}, 
Kitaev et al \cite{kitaev2002classical}, Benenti et al \cite{benenti2004principles} and Popescu et al \cite{lo1998introduction}.

The circuit model of quantum computation is based on the notion of \important{qubit}. A qubit is a physical system whose states are described by unit vectors in a two-dimensional Hilbert space
$\mathcal{H} \simeq \mathbbm{C}^2$. A system of $n$ qubits, also called an $n$-qubit \important{register}, has states described by unit vectors in the Hilbert space $\mathcal{H}_n = 
\mathcal{H}^{\otimes n}$. An orthonormal basis of the Hilbert space $\mathcal{H}_n$ is given by the following vectors, called \important{computational basis states},
\begin{equation}
\label{eq:qcomputational_basis}
\ket{ {\bf{z}} } 
= 
\bigotimes_{\ell=0}^{n-1} \ket{ z_\ell }
=
\ket{ z_{n-1} \dots z_0 } 
= 
\ket{ z } 
\quad,\quad
{\bf{z}} \in \{0,1\}^n
\quad,\quad
z = \sum_{\ell=0}^{n-1} z_\ell \, 2^\ell \in \{0 \dots 2^n-1 \}
\quad.
\end{equation}
Starting from a register of $n$ qubits prepared in the state $\ket{0} \in \mathcal{H}_n$, a generic $n$-qubit state $\ket{\psi}$ can be prepared applying single- and multi-qubit 
unitary transformations, or \important{gates}. 
Examples of single-qubit and two-qubit gates are listed in Table \ref{table:gates}. In quantum computation, a role of particular importance is played by \important{Pauli operators}, defined as
\begin{equation}
\hat{\sigma}_{{\bf{m}}} = \hat{\sigma}_{m_{n-1}} \otimes \dots \otimes \hat{\sigma}_{m_0}
\quad,\quad
\hat{\sigma}_m \in \{ \mathbbm{1} , X, Y, Z \}
\quad,
\end{equation}
where the single-qubit Pauli operators are illustrated in Table \ref{table:gates}.

\begin{table}[b!]
\includegraphics[width=\textwidth]{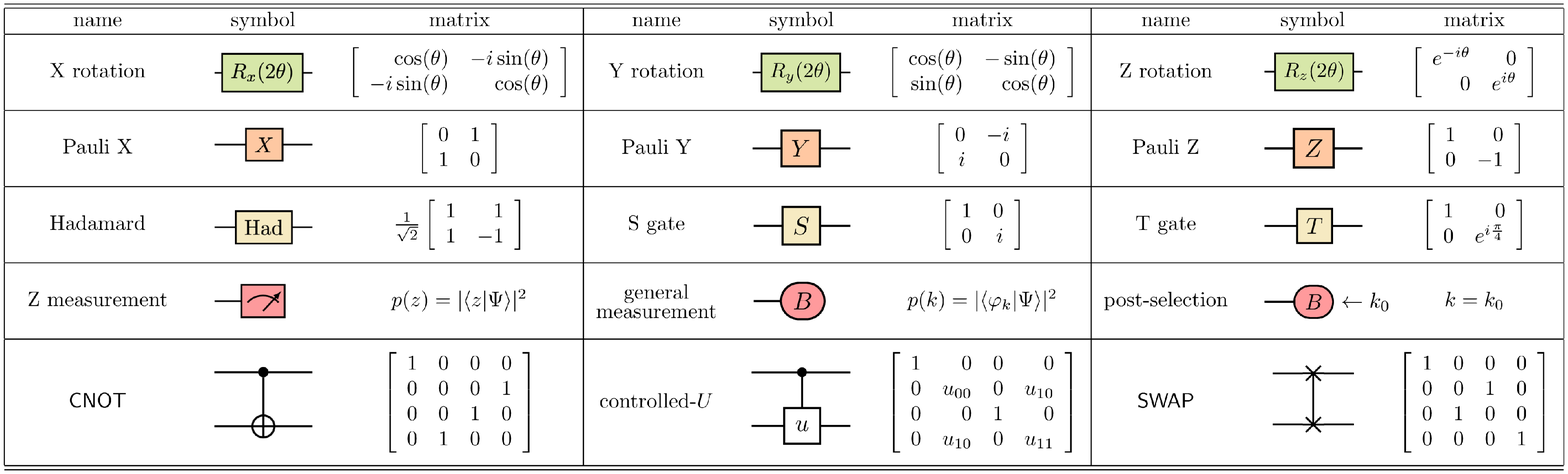}
\caption{Examples of quantum gates and circuit elements. From top to bottom: single-qubit rotations (i.e. exponentials of single-qubit Pauli operators), for example $R_x(\theta) = \exp(-i \theta X/2)$. Single-qubit Pauli operators, which are equal to special single-qubit Pauli rotations up to a global phase, for example $X = R_x(\pi/2)$.
Single-qubit operations in the Clifford group (Hadamard, $S$ and $T$ gates), which are equal to special single-qubit Pauli rotations up to a global phase, namely $S = R_z(\pi/2)$, $T=R_z(\pi/4)$ 
and $H = \exp(-i \pi/2 (X+Z))$.
Measurement of a single qubit in the computational basis,
measurement of an observable $B = \sum_k b_k | \varphi_k \rangle \langle \varphi_k |$, and measurement of an observable with post-selection (retaining only one specific outcome $k_0$).
Two-qubit $\mathsf{CNOT}$ (controlled-$X$), $\mathsf{cU}$ (controlled-$U$) and $\mathsf{SWAP}$ gates. The $\mathsf{CNOT}$ gate is sometimes denoted
$\mathsf{CNOT}_{ij}$, where $i$ and $j$ are called 
the control and target qubit respectively, and applies an $X$ transformation to its target qubit ($\oplus$ symbol) if its control qubit ($\bullet$ symbol) is in the state $|1\rangle$, the $\mathsf{cU}$ can be written as a product of up to two $\mathsf{CNOT}$ gates and four single-qubit gates, and the $\mathsf{SWAP}$ gate can be written as a product of three $\mathsf{CNOT}$ gates, $\mathsf{SWAP}_{ij} = \mathsf{CNOT}_{ij} \mathsf{CNOT}_{ji} \mathsf{CNOT}_{ij}$. 
{Qubits are ordered from top to bottom, and matrix elements are defined as $G_{zw} = \langle z | \hat{G} | w \rangle = \langle z_{n-1} \dots z_0 | \hat{G} | w_{n-1} \dots w_0 \rangle$, with binary digits running from right to left as in Eq.~\eqref{eq:qcomputational_basis}.}
}
\label{table:gates}
\end{table}

\begin{figure}
\includegraphics[width=0.8\textwidth]{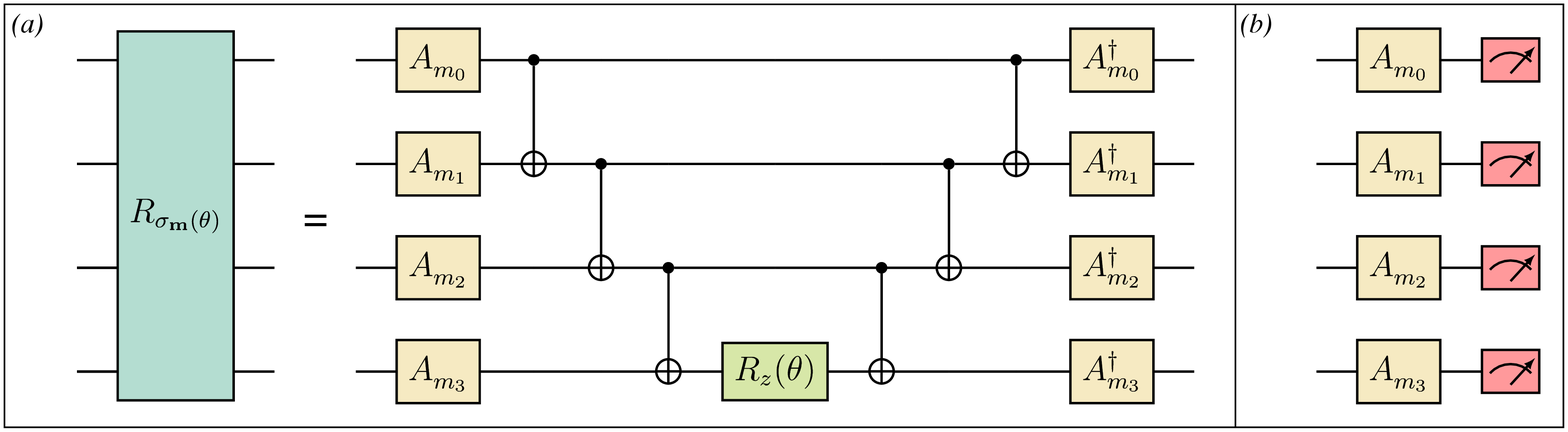}
\caption{Quantum circuits (a) to apply the exponential of a Pauli operator $\otimes_{\ell=0}^{3} \sigma_{m_\ell}$
and (b) to measure its expectation value. The single-qubit gates are {$A_X = H$, $A_Y = HS$ and $A_Z = I$}.
}
\label{fig:pauli_circuits}
\end{figure}

Pauli operators are a basis for the space of linear operators on $\mathcal{H}_n$. Exponentials of Pauli operators {can be
represented as tensor products of $Z$ operators,}
\begin{equation}
{
\hat{R}_{ \hat{\sigma}_{{\bf{m}}} }(\theta) 
= 
e^{ -\frac{i \theta}{2} \hat{\sigma}_{{\bf{m}}} } 
= 
\hat{V}^\dagger 
e^{ - \frac{i \theta}{2} Z \otimes \dots \otimes Z } \hat{V}
\quad,\quad
\hat{V} 
= 
\bigotimes_{\ell=0}^{n-1} \hat{A}_{m_\ell}
\quad,\quad
\hat{A}_{m}^\dagger 
\hat{\sigma}_m 
\hat{A}_{m}
= 
Z
\quad,
}
\end{equation}
{and then applied to a register of qubits as illustrated in Fig.~\ref{fig:pauli_circuits}a. Such a circuit contains ladders of $\CNOT$ gates that compute 
the total parity into the last qubit \cite{seeley2012bravyi}.}

Two parameters, respectively called \important{width and depth}, are often used to characterize the cost of a quantum circuit. Width refers to the number of qubits that comprise the circuit (in Fig.~\ref{fig:pauli_circuits}a, $\mathsf{width}=4$). Depth refers to the number of layers of gates that cannot be executed at the same time (in Fig.~\ref{fig:pauli_circuits}a, $\mathsf{depth}=9$). Although width and depth are both limiting factors in the execution of quantum algorithms (large width corresponds to many qubits, and large depth to many operations in presence of decoherence), the latter is a major computational bottleneck on near-term devices.
The error mitigation techniques presented in Section \ref{ref:sec_simulations} are used to increase the width and depth of circuits that can be executed on near-term hardware.

Furthermore, Pauli operators are Hermitian, and thus can be measured. 
Since
\begin{equation}
\hat{\sigma}_{{\bf{m}}} 
=
\hat{V}^\dagger {(Z \otimes \dots \otimes Z)} \hat{V}
=
\sum_{ {\bf{z}} } f({\bf{z}})
\hat{V}^\dagger | {\bf{z}} \rangle \langle {\bf{z}} | \hat{V}
\quad,\quad
f({\bf{z}}) = \prod_{\ell=0}^{n-1} (-1)^{z_\ell}
\quad,
\end{equation}
preparing a register of qubits in a state $| \Psi \rangle$, applying the transformation $\hat{V}$ and measuring all qubits 
in the computational basis as shown in Fig.~\ref{fig:pauli_circuits}b yields a collection of samples, or "shots", 
$\{ {\bf{z}}_i \}_{i=1}^{n_s}$, from which the expectation value $\langle \Psi | \hat{\sigma}_{{\bf{m}}} | \Psi \rangle$ can
be estimated as $\mu \pm \sigma$ with $\mu = n_s^{-1} \sum_i f( {\bf{z}}_i )$ and $\sigma^2 = n_s^{-1} (1-\mu^2)$.

{The presence of statistical uncertainties on measurement results is a basic but centrally important aspect of quantum computation: quantum algorithms must be understood and formulated in terms of random variables and stochastic calculus, and their results accompanied with carefully estimated statistical uncertainties. These aspects 
cannot be overlooked in the design and implementation of quantum algorithms.}

In the remainder of this work, we will call measurements and unitary transformations executed on a digital quantum computer \important{operations}, and use the term gates to denote unitary transformations only.

In this section, we provided a very concise presentation of the concepts of qubit, quantum gates and measurements,
with the purpose of fixing notation and maintaining the remainder of the work self-contained. 
In the coming sections, we will show that suitable sets of single- and two-qubit gates are universal, i.e. they can be 
multiplied to yield a generic unitary transformation, and present strengths and limitations of quantum computation in 
the light of quantum complexity theory.

\subsubsection{Universality and limitations of digital quantum computers}
\label{sec:universality}

Informally, a set $\mathcal{S}$ of quantum gates is called \important{universal}
if any unitary transformation that can be applied on a quantum computer can be 
expressed as a product of a finite number of gates from $\mathcal{S}$.
Any multi-qubit gate can be expressed as a product of single-qubit and $\CNOT$ operations only, and therefore $\mathcal{S}_1 = \{ \mbox{single-qubit gates}, \mathsf{CNOT} \}$ is a universal set of quantum gates.

Since single-qubit operations have continuous parameters (corresponding to rotation angles), the set $\mathcal{S}_1$ is not countable. However, it is known that the set $G$ of single-qubit gates generated by $\{\mathrm{\hadamard,S,T}\}$ is dense in SU(2) \cite{nielsen2002quantum}. According to the Solovay-Kitaev theorem \cite{kitaev1997quantum,nielsen2002quantum,kitaev2002classical,harrow2002efficient,dawson2005solovay}, for any $\hat{U} \in$ SU(2) and any target accuracy $\varepsilon$ there exists a sequence of $\mathcal{O}(\log^c(1/\varepsilon))$ gates from the generating set of $G$ that approximates $\hat{U}$ within accuracy $\varepsilon$ and $c \simeq 4$. {Subsequent work has demonstrated \cite{kliuchnikov2012fast,ross2014optimal}
that $z$ rotations and generic single-qubit gates can be implemented with no more than $3 \, \log(1/\varepsilon)$ and $9 \, \log(1/\varepsilon)$ T and H gates respectively, which is asymptotically optimal, with a practical algorithm.
Therefore, the countable set $\mathcal{S}_2 = \{ \mathrm{\hadamard, S, T, \CNOT} \}$ is a universal set of quantum gates.}

It should be noted, however, that a generic $n$-qubit unitary 
transformation is exactly represented with {$\mathcal{O}(4^n)$} single-qubit 
and $\CNOT$ gates \cite{barenco1995elementary,mottonen2004quantum,shende2006synthesis}, meaning that quantum 
computers are not guaranteed to give access to a generic $n$-qubit state 
with an amount of operations scaling polynomially with $n$. 
Furthermore, finding an \textbf{efficient} 
decomposition of a unitary transformation in terms of elementary quantum operations can itself be a challenging task \cite{daskin2011decomposition}.

Finally, the set $\mathcal{C} = \{ \mathrm{Had, S, \CNOT} \}$ generates the so-called \important{Clifford group} of unitary transformations, that map 
Pauli operators onto Pauli operators. An important theoretical result, the Gottesman-Knill theorem \cite{gottesman1998heisenberg,aaronson2004improved,nest2008classical}, 
states that quantum circuits using only $(a)$ preparation of qubits in computational basis states, $(b)$ application quantum gates from the Clifford group and $(c)$ measurement {of a single Pauli operator}, can be efficiently simulated on a classical computer.
Therefore, the entanglement that can be achieved with circuits of 
Clifford gates alone does not give any computational advantage over 
classical computers. Computational advantage is to be sought in circuits
also containing $T$ gates.

\subsubsection{Quantum computational complexity}
\label{sec:complexity}

Solving a problem on a computational platform requires designing an algorithm, by which term we mean the 
application of a sequence of mathematical steps.
Executing an algorithm requires a certain amount of resources, typically understood in terms of space (or memory) 
and time (or elementary operations).
Computational complexity theory groups computational problems in complexity classes defined by their resource 
usage, and relates these classes to each other.
Quantum computational complexity theory is a branch of computational complexity theory that, loosely speaking, 
studies the 
difficulty
of solving computational problems on quantum computers, 
formulates quantum complexity classes, and relates such quantum complexity classes with their classical 
counterparts \cite{bernstein1997quantum,kitaev2002classical,watrous2008quantum}.
Two important quantum complexity classes are BQP (bounded-error quantum polynomial time) and QMA (quantum 
Merlin-Arthur).
Roughly speaking, BQP comprises problems that can be solved with polynomial space and time resources on a quantum computer. 
In the context of quantum simulation for quantum chemistry, the most important BQP problem is the simulation of 
\important{Hamiltonian dynamics} \cite{feynman1982simulating,feynman1986quantum,lloyd1996universal,zalka1998efficient}.
QMA, on the other hand, comprises problems where putative solutions can be verified but not computed in polynomial time on a quantum computer. Producing a putative solution means executing a quantum circuit giving access to a wavefunction $\Psi$, and verifying a putative solution means
executing a second quantum circuit to ensure that $\Psi$ is actually a solution of the problem of interest.
In the context of quantum simulation for quantum chemistry, 
the most important QMA problem is the simulation of 
\important{Hamiltonian eigenstates}, as discussed for example in \cite{kitaev2002classical,kempe2006complexity}.

Current knowledge indicates that both the ground-state and the Hamiltonian dynamics problems are, in a worst-case scenario, exponentially expensive 
on a classical computer. The Hamiltonian dynamics
problem is thus a relevant application for quantum algorithms, as it offers theoretical opportunities for better computational performance when tackled on a quantum computer.
However, there are a number of practical considerations to take into account.
 For example, the statements of computational complexity theory refer to exact solutions of the problem at hand, and experience indicates that approximate methods can deliver accurate results for certain problems, which calls
for a systematic characterization of quantum algorithms for chemistry in both accuracy and computational cost across a variety of chemical problems. Furthermore, quantum hardware has to reach a level of control and predictability compatible with large-scale quantum chemical simulations, which is one of the most important challenges confronting experimental physics and engineering.

The division between BQP and QMA problems will inform the remainder of the present work: in the next Section we will present some important problems in computational chemistry, highlighting those based on the simulation of Hamiltonian dynamics. We will then present quantum algorithms for the simulation of Hamiltonian dynamics, and heuristic algorithms for Hamiltonian eigenstate approximation.

\section{Some important problems in computational chemistry}

In this Section, we briefly review some problems studied in computational chemistry. While their extensive description is beyond the scope of the present review, 
we will attempt to highlight some technical challenges, through the lens of which the quantum algorithms presented in the forthcoming sections can be examined.

\subsection{Electronic structure}
\label{sec:es}

The main objective of electronic structure in chemistry is to determine the ground and low-lying excited states of a 
system of interacting electrons. Often relativistic effects and the coupling between the dynamics of electrons and nuclei 
can be neglected, or treated separately. Within this approximation, the many-electron wavefunction can be found by 
solving the time-independent Schr\"{o}dinger equation for the Born-Oppenheimer Hamiltonian \cite{Born_1927,Szabo_book_1989}
{
\begin{equation}
\label{eq:bo_ham}
\hat{H} \Psi
= \left[ \sum_{a<b}^{N_n} \frac{Z_a Z_b}{| \vett{R}_a - \vett{R}_b|} \right] \Psi
- \frac{1}{2} \, \sum_{i=1}^N \frac{\partial^2 \Psi}{\partial \vett{r}_i^2} 
- \left[ \sum_{i=1}^N \sum_{a=1}^{N_n} \frac{Z_a}{| \vett{r}_i - \vett{R}_a |} \right] \Psi
+ \left[ \sum_{i<j}^N \frac{1}{|\vett{r}_i - \vett{r}_j|} \right] \Psi
 \,,
\end{equation}
}
where $\vett{r}_i$ is the position of electron $i$, and $\vett{R}_a$ the position of nucleus $a$, {having atomic number $Z_a$}. 
The numbers of electrons and nuclei are $N$ and $N_n$, respectively,
nuclear positions are held fixed, and atomic units are used throughout.

Solving the time-independent Schr\"{o}dinger equation is 
needed to access molecular properties including, and not
limited to, energy differences 
(e.g. ionization potentials, electron affinities, singlet-triplet gaps, binding energies, 
deprotonation energies),
energy gradients 
(e.g. forces and frequency-independent polarizabilities)
and electrostatic properties
(e.g. multipole moments and molecular electrostatic potentials).

The first step of any electronic structure simulation is to approximate Eq.~\eqref{eq:bo_ham} with a simpler Hamiltonian
acting on a finite-dimensional Hilbert space. This is usually achieved by truncating the Hilbert space of a single electron 
to a finite set of orthonormal basis functions or spin-orbitals $\{ \varphi_p \}_{p=1}^M$. 
Electronic structure simulations based on the first quantization formalism describe a system of $N$ electrons in $M$
spin-orbitals using the configuration interaction (CI) representation
\begin{equation}
\label{eq:ci_wfn}
| \Psi \rangle = \sum_{i_1 < \dots < i_N}^M \psi_{i_1 \dots i_N} \, | i_1 \dots i_N \rangle  
\quad,
\end{equation}
where $| i_1 \dots i_N \rangle$ is the Slater determinant where orbitals $i_1 \dots i_N$ are occupied. 

Electronic structure simulations based on the second quantization formalism, on the other hand, operate in the Fock 
space of electrons in $M$ spin-orbitals, and represent the Hamiltonian as 
\begin{equation}
\label{eq:hamiltonian_born_oppenheimer}
\hat{H} 
= \sum_{a<b}^{N_n} \frac{Z_a Z_b}{| \vett{R}_a - \vett{R}_b|} 
+ \sum_{pr} h_{pr} \, \crt{p} \dst{r} 
+ \sum_{prqs} \frac{(pr|qs)}{2} \, \crt{p} \crt{q} \dst{s} \dst{r} 
\,,
\end{equation}
where $\crt{p},\dst{r}$ are fermionic creation and annihilation operators associated to spin-orbitals $\varphi_p, \varphi_r$
respectively. 

\subsubsection{Basis sets}

Choosing a finite set of spin-orbitals $\{ \varphi_p \}_{p=1}^M$ balances two concerns: the need of representing electronic states and operators
with as few
orbitals as possible, and the need for obtaining accurate results.
Gaussian bases are most commonly employed in molecular simulations due to their compactness, 
while plane waves are most often used in the simulation of crystalline solids.
In order to get quantitatively accurate results, very large one-electron basis sets have to be used, in either case.

Since, as we shall see, the number $M$ of orbitals translates 
into the number of qubits needed to perform a simulation (see
Section \ref{sec:fermions_second}) and near-term hardware 
is limited in the number and quality of qubits (see Section \ref{sec:hardware}), most quantum computing simulations so far 
have been limited to $M \simeq 10$ orbitals. Theoretical and computational chemists can help the field of quantum computation by integrating techniques (e.g. optimized orbitals \cite{mizukami2020orbital}, perturbative treatments \cite{takeshita2020increasing}, explicit electronic correlation \cite{motta2020quantum}) to account for more orbitals without extra qubits and gates.

Furthermore, most quantum algorithms have so far been tested 
and characterized using sets of $M \simeq 10$ orbitals (corresponding to small active spaces or minimal bases). While the results of such investigations have often indicated high accuracy when compared to classical methods with the same basis sets, it is in many cases uncertain whether known algorithms are useful when larger bases are used.
Thus, another area of research where theoretical and computational chemists can offer insight and help is the systematic extension and benchmark of quantum algorithms beyond the small basis sets investigated so far.

\subsubsection{Classical algorithms and open problems}

The main obstacle to the investigation of electronic structure is that, in general, the computational cost of finding the 
exact eigenfunctions of Eq.~\eqref{eq:bo_ham} grows combinatorially with the size of the studied system \cite{Troyer_PRL_2004,Schuch_NAT_2009}.
This limitation has so far precluded exact studies for all but the smallest systems and motivated the development of approximate methods.
At a high level, those methods can be distinguished by general categories such as wavefunction, embedding, and diagrammatic {(or Green's function)}.

\paragraph{Wavefunction methods} 
formulate an Ansatz for an eigenstate, e.g. the ground state, and compute expectation values of observables 
and correlation functions with respect to that wavefunction. The nature of the underlying Ansatz is ultimately responsible for the accuracy and computational cost of a given method.  For molecular systems, a hierarchy of quantum chemistry methods has been developed, which allow systematic improvement in accuracy, at increasing computational cost. 
These techniques typically have their starting point in the Hartree-Fock (HF) method, which approximates the ground state of a molecular Hamiltonian with the lowest-energy Slater determinant, and incorporate electronic correlation.
For example, one of the most accurate methods is coupled-cluster with singles and doubles and perturbative estimate to the connected triples, CCSD(T) \cite{Paldus_ACP_1999,bartlett2007coupled,Shavitt_book_2009}. 
Other promising alternative approaches include tensor network methods \cite{White_PRL_1992,White_JCP_1999,Chan_JCP_2002,Olivares_JCP_2015,Chan_JCP_2016}, 
which represent electronic wavefunctions as contractions between tensors, 
and quantum Monte Carlo methods \cite{Booth_JCP_2009,Booth_NAT_2013,Reynolds_JCP_1982,Foulkes_RMP_2001,Zhang_PRL90_2003,motta2018ab,motta2017towards}, 
which instead represent electronic properties as expectation values of carefully designed random variables.

\paragraph {Density functional (DFT) methods }
The most widely used 
methods are 
mean field in nature and are based on the density \cite{Martin_book_2004,Kohn_RMP_1999}, making them less expensive than wavefunction methods. DFT methods are based on an approximation to the Hamiltonian rather than an approximation to the wavefunction and there is no hierarchy of functionals as there is in wavefunction methods \cite{hammes2017conundrum}.
Nevertheless, density functional methods are standard tools for electronic structure calculations in many areas across multiple disciplines, with sophisticated computer software packages available due to their significantly lower cost. 

\paragraph{Embedding methods}
evaluate the properties of a large system by partitioning it within a given basis (e.g. the spatial or energy basis) into a collection of fragments, 
embedded in a self-consistently determined environment
\cite{Knizia-2012,Knizia-2013,Wouters-2016,Georges1996dynamical,vollhardt2012dynamical}. 
These methods combine two different types of quantum calculations: high-level calculations on fragments,
and low-level calculations on the environment surrounding fragments.
The accuracy of an embedding method is determined by a combination of several factors including: 
the size of the fragments, the accuracy in the treatment of the embedded fragments and environment, and the convergence of the self-consistency 
feedback loops between fragments and their environment. 

\paragraph{Diagrammatic methods}
evaluate, either deterministically or stochastically, a subset of the terms in the diagrammatic interaction expansion
\cite{Hedin,VanHoucke2010,Kulagin2013,nguyen2016rigorous,AlexeiSEET,LAN} {of a quantity such as the Green's function, the self-energy, or the ground-state energy}.
These methods are often based on the Feynman diagrammatic technique formulated in terms of self-consistent propagators and bare or renormalized interactions, at finite or zero temperature,
and their accuracy and computational cost is determined by several factors, including the subsets of diagrams or series terms that are included in the calculation. 

\begin{figure}[b!]
\centering
\includegraphics[width=0.75\textwidth]{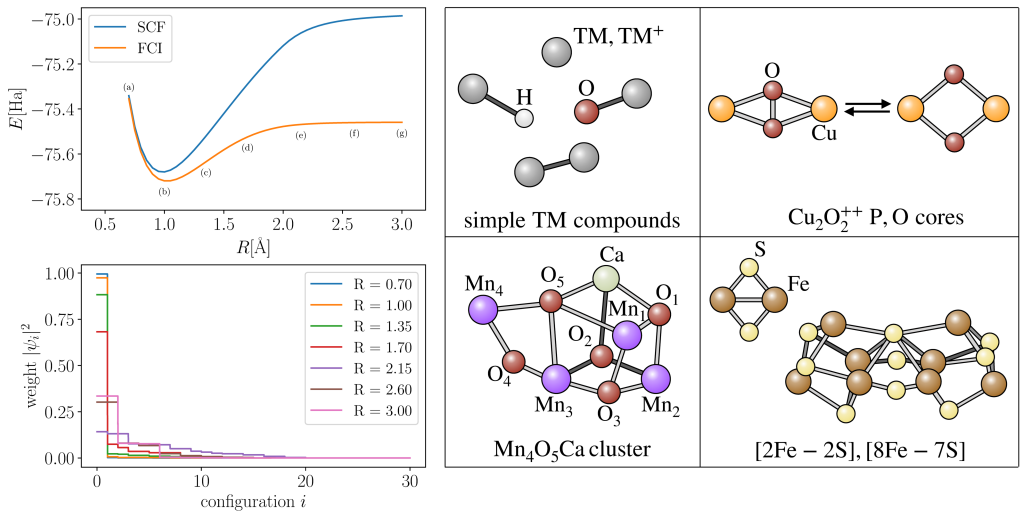}
\caption{Left: example of multi-reference character in the dissociation of H$_2$O. Right: transition metal (TM) compounds
of relevance for fundamental chemistry and quantum computing research: TM atoms, ions, hydrides and oxides can constitute subjects of study for near-term quantum devices. Longer-term goals are the description of CuO, MnO and FeS cores found in a variety of biological catalysts.}
\label{fig:es}
\end{figure}

These methodologies tend to be most accurate for problems where a single electronic configuration dominates the representation Eq.~\eqref{eq:ci_wfn} of the ground-state wavefunction,
the so-called single-reference problems, exemplified by the ground states of many simple molecules at equilibrium.
In many molecular excited states, along bond stretching, and in transition metal systems, 
multiple electronic configurations contribute to the ground-state wavefunction, as illustrated
in Fig.~\eqref{fig:es}a,
leading to multi-reference quantum chemistry problems.

Despite remarkable progress in the extension of quantum chemistry methods to multi-reference situations,
the accuracy attainable for molecules with more than a few atoms is considerably lower than in the single-reference case.
Some important examples of multi-reference quantum chemical problems are sketched in Fig.~\eqref{fig:es}b.
Transition metal atoms, oxides, and dimers pose formidable challenges to even remarkably accurate many-body methods, 
as they feature multiple bonds, each with a weak binding energy, and potential energy surfaces resulting from the interplay between \important{static and dynamical} electronic correlation
\cite{williams2020direct,AlSaidi_PRB73_2006,Purwanto_JCP142_2015,Purwanto_JCP144_2016,shee2019achieving,shee2021revealing}. Full configuration interaction (FCI) methods can describe such an interplay but systems that can be studied with FCI methods are arguably small \cite{rossi1999full,vogiatzis2017pushing}.
Coupled binuclear copper centers are present in the active sites of some very common metalloenzymes found in living organisms.
The correct theoretical description of the interconversion between the two dominant structural isomers of the $\ce{Cu2O2^{2+}}$ core 
is key to the ability of these metalloenzymes to reversibly bind to molecular oxygen,
and is challenging since it also requires a method that can provide a balanced description of static and dynamic electronic correlations
\cite{solomon1996multicopper,kitajima1994copper,samanta2012exploring,gerdemann2002crystal}.

The chemistry of other active sites, such as the $\ce{Mn3CaO4}$ cubane core of the oxygen-evolving complex of photosystem II,
and the $\ce{Fe7MoS9}$ cofactor of nitrogenase, 
pose some of the most intricate multi-reference problems in the field of biochemistry.
In particular with respect to their spectroscopic properties, 
such systems require
a detailed characterization of the interplay between spin-coupling and electron delocalization between metal centers.
\cite{sharma2014low,kurashige2013entangled,li2019electronic,li2019electronic2,chilkuri2019ligand,cao2018protonation}.

\subsection{Electronic dynamics}
\label{sec:electron_dynamics}

Many experiments conducted on molecules probe their dynamical, rather than equilibrium, properties. An important example are
oscillator strengths
\begin{equation}
D_{0\to f} \propto (E_f - E_0) \sum_{\alpha=x,y,z} | \langle \Psi_f | \hat{\mu}_\alpha | \Psi_0 \rangle |^2
\quad,
\end{equation}
where $\Psi_0$ and $\Psi_f$ are a ground and excited state of the electronic Hamiltonian, with energies $E_0$ and $E_f$ respectively, and $\hat{\mu}_\alpha$ is the dipole moment along direction $\alpha=x,y,z$. Oscillator strengths exemplify properties, such as structure factors \cite{damascelli2003angle,Damascelli_2004},
that are determined by excited electronic states through 
transition energies and matrix elements.
A variety of algorithmic tools are available for the calculation of spectral functions and time-dependent properties on classical computers today, including
time-dependent density functional theory, equation-of-motion coupled-cluster and diagrammatic theories \cite{onida2002electronic,stanton1993equation}.
In some regimes (e.g. transitions to specific and structured excited states) dynamical properties such as oscillator strengths can be computed efficiently and accurately. On the other hand, there are computationally challenging regimes, where quantum algorithms can be relevant (e.g. congested spectra) owing to its ability to simulate time evolution.

In fact, oscillator strengths are straightforwardly obtained from dipole-dipole \important{structure factors}
\begin{equation}
S_{\mu_\alpha,\mu_\beta}(\omega) 
= 
\sum_l 
\langle \Psi_0 | \hat{\mu}_\alpha | \Psi_l \rangle
\, 
\delta\Big( \hbar \omega - (E_l-E_0) \Big) 
\,
\langle \Psi_l | \hat{\mu}_\beta  | \Psi_0 \rangle 
\quad,
\end{equation}
as $D_{0\to f} = \sum_\alpha S_{\mu_\alpha,\mu_\alpha}(E_f-E_0)$. Dipole-dipole structure factors are in turn the Fourier transform of the time-dependent
dipole-dipole \important{correlation functions},
\begin{equation}
\label{eq:linear_response1}
C_{\mu_\alpha,\mu_\beta}(t) 
= 
\langle \Psi_0 | \hat{\mu}_\alpha 
\,
e^{ \frac{t}{i\hbar} (\hat{H}-E_0)}
\,
\hat{\mu}_\beta | \Psi_0 \rangle
=
\sum_f 
\langle \Psi_0 | \hat{\mu}_\alpha | \Psi_f \rangle
\,
e^{ \frac{t}{i\hbar} (E_f-E_0)}
\,
\langle \Psi_f | \hat{\mu}_\beta  | \Psi_0 \rangle 
\quad,
\end{equation}
and can be computed by simulating time evolution.

Another important and related research theme is the 
computation of \important{time- and frequency-dependent properties} 
(e.g. non-linear optical properties, electro-optical effects, circular dichroism).
Similar to oscillator strengths, these properties 
are challenging to compute as they involve excited states,
and are natural applications for quantum algorithms,
as they can be computed by simulating time evolution.

Consider a system at equilibrium in the ground state $\Psi_0$ of $\hat{H}_S$ at time $t=0$, and subject to a perturbation of the 
form $\hat{V}(t) = \sum_k f_k(t) \, \hat{O}^\dagger_k$. The expectation value of the operator $\hat{O}_j$ at time $t>0$ is given by
\begin{equation}
O_j(t) = \langle \Psi(t) | \hat{O}_{j,S}(t) | \Psi(t) \rangle
\quad,\quad
| \Psi(t) \rangle = \hat{U}(t) | \Psi_0 \rangle
\quad,\quad
\hat{U}(t) = \sum_{n=0}^{\infty} \int_0^t \frac{dt_1}{i\hbar} \dots \, \int_0^{{t_{n-1}}} \frac{dt_n}{i\hbar} \, \hat{V}_S(t_1) \dots \hat{V}_S(t_n)
\quad,
\end{equation}
where $\hat{V}_S(t) = e^{ - \frac{t \hat{H}_S}{i \hbar} } \hat{V} e^{ \frac{t \hat{H}_S}{i \hbar} }$. 
The linear response approximation \cite{fetter2012quantum}, 
valid for weak external perturbations, {consists of} truncating $\hat{U}(t)$ and $O_j(t)$ to first order in $\hat{V}(t)$, leading to the expression
\begin{equation}
\label{eq:linear_response2}
O_j(t) = O_j(0) + \int_0^t \frac{dt^\prime}{i \hbar} \, \sum_k f_k(t) \, \alpha_{jk}(t-t^\prime)
\quad,\quad
\alpha_{jk}(t) = \langle \Psi_0 | [ \hat{O}_{j,S}(t) , \hat{O}^\dagger_k ] | \Psi_0 \rangle 
\quad,
\end{equation}
where $\alpha_{jk}(t)$
is the time-dependent 
polarizability of $\hat{O}_j$, $\hat{O}_k$.

Real-time correlation functions such as \eqref{eq:linear_response1} and \eqref{eq:linear_response2} 
are objects of central importance 
in many-particle physics, but naturally emerge in the 
framework of linear response theory, i.e. of weak external perturbation. Going beyond spectral properties, and probing 
the non-equilibrium real-time dynamics of chemical systems, 
is a research topic of increasing relevance, both because of experiments that can now probe quantum dynamics at atomic scales, and because of fundamental interest in studying the time-dependent behavior of many-particle systems.

\subsection{Molecular vibrations}

The geometric structure of molecules is typically probed by gas-phase 
spectroscopic experiments, whose interpretation,
for all but the smallest molecules, needs input from numerical simulations, due to many competing transitions between states
\cite{barone2015quantum,bowman2008variational,X3}.
Numerical simulations of molecular rovibrational levels {require} solving the Schr\"{o}dinger equation for the nuclei. 
Within the Born-Oppenheimer approximation, the nuclear Hamiltonian has the form
\begin{equation}
\label{eq:nuclear_bo}
\hat{H}_{nuc} = - \frac{1}{2} \sum_{a=1}^{N_n} \frac{1}{M_a} \frac{\partial^2}{\partial \vett{R}_a^2} + V( \vett{R}_1 \dots \vett{R}_{N_n} )
 \,,
\end{equation}
where $V$ denotes the ground-state potential energy surface of the electronic Hamiltonian Eq.~\eqref{eq:bo_ham} when nuclei have positions 
$\{ \vett{R}_a \}_{a=1}^{N_n}$, and $\{ M_a \}_{a=1}^{N_n}$ are the nuclear masses, in atomic units.
Solving the nuclear Schr\"{o}dinger equation presents additional challenges: first of all, as the interaction among nuclei is mediated by
electrons, the function $V$ is not known a priori, and needs to be computed from quantum chemical calculations at fixed nuclear 
geometries as in Section \ref{sec:es}, and then fitted to an appropriate functional form, which can be an expensive procedure \cite{X1,X2}.
For certain problems, the harmonic oscillator approximation of $V$ is appropriate, which is given by \cite{X1}
\begin{equation}
\label{eq:nuclear_bo_harmonic}
\hat{H}_{nuc} = \frac{1}{2} \sum_{\alpha\beta} (\hat{J}_\alpha - \hat{\pi}_\alpha) \mu_{\alpha\beta} (\hat{J}_\beta - \hat{\pi}_\beta) - \frac{1}{2} \sum_k \frac{\partial^2}{\partial Q_k^2} - \frac{1}{8} \sum_\alpha \mu_{\alpha\alpha}
+V({\bf{Q}})
\quad,\quad
V({\bf{Q}}) = \frac{\kappa}{2} {\bf{Q}}^2
\quad.
\end{equation}
In Eq.~\eqref{eq:nuclear_bo_harmonic} $\hat{J}_\alpha$ is the total angular momentum of a given cardinal direction ($x$, $y$, or $z$) denoted by $\alpha$ or $\beta$; $\hat{\pi}_\alpha$ is the total vibrational angular momentum of the same direction; $\mu_{\alpha\beta}$ $\mu_{\alpha\alpha}$ is the inverse of the moment of inertia tensor for the given geometric coordinate; $Q_k$ is a single normal coordinate, and ${\bf{Q}}$ is the set of all normal coordinates; and $\kappa$ is a numerically determined spring constant.

Notwithstanding its simplicity (and usefulness in situations such as correction of binding energies for zero-point nuclear motion),
the harmonic approximation Eq.~\eqref{eq:nuclear_bo_harmonic} has several limitations. 
As a result of equal spacing of energy levels for a given normal mode, all transitions occur at the same frequency, and bond dissociation is not described.
Improving over the harmonic approximation requires retaining higher-order or anharmonic terms in $V$, 
which poses additional challenges, especially a proper choice of curvilinear nuclear coordinates,
and developing accurate solvers for the nuclear Schr\"{o}dinger equation \cite{sadri2012numeric,tew2003internal,viel2017zeropoint,beck2001multiconfiguration,X1,X2}.

Computing solutions of the nuclear Schr\"{o}dinger equation has many valuable applications,  notably supporting the study of chemical processes  occurring in atmospheric chemistry,  and the identification of molecules in the interstellar medium and circumstellar envelopes
\cite{vaida2008spectroscopy,heiter2015atomic,smith1988formation,schilke2001line,agundez2008tentative,agundez2014new,X1,X2}.

\subsection{Chemical reactions}

Understanding the microscopic mechanisms underlying chemical reactions is another problem of central importance in chemistry.
A particularly important goal is to calculate thermochemical quantities, such as reactant-product enthalpy and free energy differences, and activation energies.

A somewhat common misconception in the quantum
computation community stems from the incorrect 
interpretation and use of the term "chemical accuracy".
As the Arrhenius equation postulated that temperature dependence of the reaction rate constants is contained in an exponential factor
of the form $\exp( - \beta E_a)$, where $E_a$ is the activation energy, computed reaction rates differ by their experimental values 
by an order of magnitude at room temperature, $\beta^{-1} \simeq 0.5922$ kcal/mol, when the energy difference $E_a$ is biased by 
$\Delta E \simeq 1.36$ kcal/mol. In the light of these consideration, the accuracy required to make realistic thermochemical predictions,
called \important{chemical accuracy}, is generally 
considered to be 1 kcal/mol \cite{pople1999nobel}. 
The term chemical accuracy refers to agreement 
between computed and experimental energy differences
within 1 kcal/mol. 
Part of the quantum computing literature
has used the term chemical accuracy to indicate agreement between computed and exact (i.e. FCI) total energies in a fixed basis set to within 1 kcal/mol, which we argue should instead be called \important{algorithmic accuracy}.

Determining the main features of the potential energy surface, i.e. the electronic energy as a function of nuclear positions, 
its minima and saddle points, is key to understanding chemical reactivity, product distributions, and reaction rates.
On the other hand, several other factors affect chemical reactions. Potential energy surfaces are shaped by the presence
and properties of solvents: indeed, 
it is 
known that solvents have the ability to modify the electron density, stabilizing transition states and intermediates and  lowering activation barriers, e.g.
\cite{hartshorn1973aliphatic,wade2006organic}.

The systematic incorporation of \important{solvation effects}, within a hierarchy of implicit \cite {cramer1999implicit,klamt2011cosmo}, hybrid QM/MM where the solvent is treated with a force field (MM) \cite {mennucci2012polarizable,senn2009qm}, and other multi-scale methods including embedding methods QM(active)/QM(surrounding) where the QM(active) refers to treatment of the most relevant part of the system with higher level quantum mechanics and QM(surrounding) refers to treatment of the surrounding by a lower level of quantum mechanics \cite{cramer1999implicit,klamt2011cosmo,mennucci2012polarizable,senn2009qm}, 
is thus necessary to improve the description of chemical systems beyond gas-phase properties, and is especially important in the description of  many chemical reactions, 
such as $\mathrm{S_N2}$ nucleophilic substitution reactions \cite{hartshorn1973aliphatic,wade2006organic}.

Conformational effects are another important aspect of 
chemical reactivity, especially in molecular crystals.
Molecular crystals have diverse applications in fields 
such as pharmaceuticals and organic electronics. 
The organic molecules comprising such crystals are bound 
by weak dispersion interactions, and as a result the same molecule may crystallize in several different solid forms, known as \important{polymorphs} \cite{bernstein2020polymorphism,price2016can,day2007strategy}.
The energy differences between polymorphs are typically within 0.5 kcal/mol or less and the structural differences between
polymorphs govern their physical properties and  functionality.
These observations call for the accuracy of first-principles quantum mechanical approaches, 
the use and refinement of many-body dispersion methods,
and efficient configuration space exploration,
making the computational characterization of molecular crystals one of the most difficult and yet highly important in
molecular chemistry \cite{bernstein2020polymorphism,lombardo2017silico,kamat2020diabat,beran2016modeling}. 

\section {Quantum algorithms for chemistry}
\label{sec:algo}

\subsection{Mappings to qubits}
\label{sec:mapping_to_qubits}

In this Section, techniques to map fermionic and bosonic degrees of freedom to qubits are presented.
The second quantization formalism for fermions is discussed in Section \ref{sec:fermions_second},
its counterpart for bosons in Section \ref{sec:bosons_second},
and alternative approaches are briefly reviewed in Section \ref{sec:fermions_bosons_alternatives}.
 
\subsubsection{Fermions in second quantization}
\label{sec:fermions_second}

The Fock space $\mathcal{F}_{-}$ of fermions occupying $M$ spin orbitals has the same dimension, $2^M$,
of the Hilbert space of $M$ qubits, $\mathcal{H}^{\otimes M}$. Therefore, it is possible to construct {a one-to-one correspondence}
\begin{equation}
{\hat{F}} : \mathcal{F}_{-} \to \mathcal{H}^{\otimes M} 
\;,\;
| \Psi \rangle 
\mapsto {\hat{F}} | \Psi \rangle \equiv | \Psi^\prime \rangle
\end{equation}
to represent fermionic wavefunctions $| \Psi \rangle$ and operators $\hat{A}$ by qubit 
wavefunctions $| \Psi^\prime \rangle$ and operators $\hat{B}^\prime = {\hat{F}} \hat{B} {\hat{F}}^{-1}$
as in Fig.~\ref{fig:simulation}.
There are combinatorially many ways to map a quantum system to a set of qubits \cite{wu2002qubits,batista2004algebraic} and,
since fermions exhibit non-locality of their state space, due to their antisymmetric exchange statistics, any representation 
of fermionic systems on collections of qubits must introduce non-local structures \cite{bravyi2002fermionic}.

\begin{figure}[b!]
\includegraphics[width=0.88\textwidth]{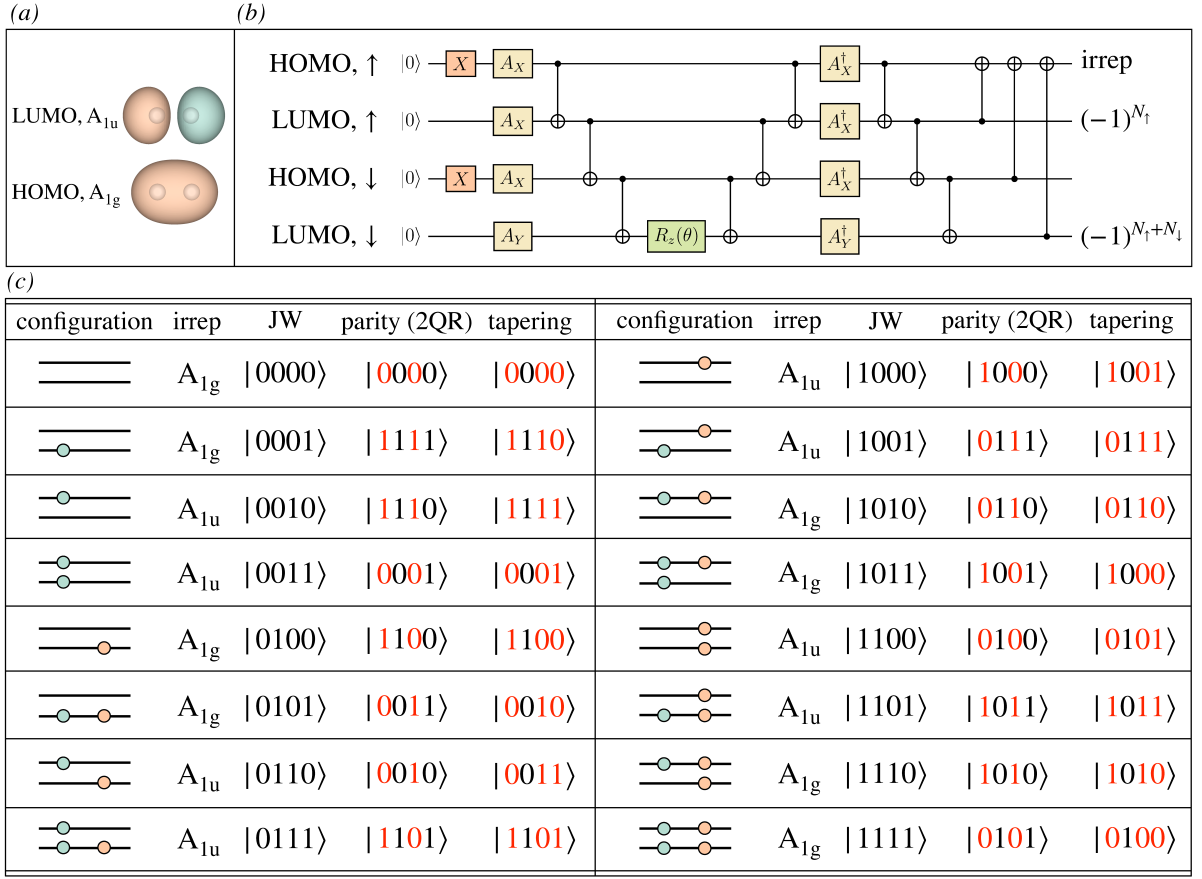}
\caption{Top: (a) molecular orbitals of the $\ce{H2}$ molecule at equilibrium geometry using a minimal STO-6G basis;
(b) quantum circuit to prepare the Hartree-Fock state ($X$ gates)
and a superposition of the Hartree-Fock and a doubly excited state
$\cos(\theta) |0101\rangle + \sin(\theta) |1010\rangle$ (exponential of the {$YXXX$} operator) in JW representation, to convert from JW to parity representation (subsequent 3 $\mathsf{CNOT}$ gates), and to compute the total irrep of the wavefunction (last 3 $\mathsf{CNOT}$ gates);
(c) mapping of electronic configurations (horizontal black lines denote molecular orbitals and blue, orange circles denote spin-up, spin-down particles)
in Jordan-Wigner representation and parity representation, and 
with tapering of the D$_{2h}$ symmetry group. Red digits denote (left to right) $(-1)^{N_\uparrow + N_\downarrow}$, $(-1)^{N_\uparrow}$, and the total irrep of the electronic configuration, $x_1+x_3 = p_0 + p_1 + p_2 + p_3$ mod 2.}
\label{fig:representations}
\end{figure}

\paragraph{Jordan-Wigner (JW) transformation.} The JW transformation
\cite{jordan1993paulische,abrams1997simulation,ortiz2001quantum,somma2002simulating}, 
maps electronic configurations with generic particle number onto computational basis states,
\begin{equation}
\big( \crt{M-1} \big)^{x_{M-1}}
\dots
\big( \crt{0} \big)^{x_0}
| \emptyset \rangle
\mapsto 
| \vett{x} \rangle 
\;,
\end{equation} 
and fermionic creation and annihilation operators ($\crt{k}$ and $\dst{k}$ respectively)
onto non-local qubit operators of the form
\begin{equation}
\label{eq:jwgivescar}
\crt{k} \mapsto 
\frac{X_k - i Y_k}{2}
\otimes Z_{k-1} \otimes \dots \otimes Z_0
\equiv
S_{+,k} \, Z^{k-1}_0
\quad,\quad
\dst{k} \mapsto 
\frac{X_k + i Y_k}{2} \, Z^{k-1}_0
\equiv
S_{-,k} \, Z^{k-1}_0
\quad.
\end{equation} 
The non-locality of these operator is required to preserve canonical anticommutation relations between creation and destruction operators,
and immediately translates to $n$-body fermionic operators.
The main limitation of the Jordan-Wigner transformation is that, as a consequence of the non-locality of the operators $Z^b_a$, the
number of qubit operations required to simulate a fermionic operator $\crt{k}$ scales as $\mathcal{O}(M)$ \cite{aspuru2005simulated,whitfield2011simulation}. 
Such a limitation motivated the design of alternative transformations, such as the parity mapping described below.
The JW transformation is exemplified in Fig.~\ref{fig:representations} using the hydrogen molecule in a minimal basis as an example. In this work, we follow the convention of mapping spin-up and spin-down orbitals to the first and the last $M/2$ qubits respectively. It is worth emphasizing that, since the JW transformation operates in the Fock space, states with any particle number, spin, and point group symmetry can result.

\paragraph{Parity transformation.}
The operator $Z^{k-1}_0$ computes the parity $p_k = \sum_{j=0}^{k-1} x_j \, \mathsf{mod} \, 2$ 
of the number of particles occupying orbitals up to $k$.
The computation of parities can be achieved using only single-qubit $Z$ operators using the following parity transformation,
\begin{equation}
\big( \crt{M-1} \big)^{x_{M-1}}
\dots
\big( \crt{0} \big)^{x_0}
| \emptyset \rangle
\mapsto 
| \vett{p} \rangle 
\quad,\quad
\crt{k} \mapsto X_{M-1} \otimes \dots \otimes X_{k+1} \otimes
\frac{ X_k - i Y_{k}}{2}
\equiv
X^{M-1}_{k+1} \otimes S_{+,k}
\;.
\end{equation}
In the parity transformation, the calculation of parities is local
but the change of occupation numbers brought by the application of
a creation or destruction operator requires $\mathcal{O}(M)$ single-qubit gates.
The parity transformation, therefore, does not improve the efficiency over that of the JW transformation, but naturally
allows a reduction of two in the number of qubits, which is 
desirable when very small quantum computers are used 
\cite{bravyi2017tapering}.
The JW and parity transformations are exemplified in Fig.~\ref{fig:representations}b,c. In JW representation, computational basis states $|x_3,x_2,x_1,x_0 \rangle$ encode
occupations $x_k$ of spin-orbitals; in parity representation, they encode parities $|p_3,p_2,p_1,p_0 \rangle$ with $p_k = x_0 + \dots + x_k$.

\paragraph{Bravyi-Kitaev (BK) transformation.} This transformation balances the locality of the occupation numbers and that of parities,  
to achieve a mapping of fermionic creation and destruction operators onto $\mathcal{O}(\log_2 M)$-qubit operators \cite{bravyi2002fermionic}.
It does so by mapping occupation number states onto suitably defined binary strings,
\begin{equation}
\big( \crt{M-1} \big)^{x_{M-1}}
\dots
\big( \crt{0} \big)^{x_0}
\mapsto 
| \vett{b} \rangle 
\;,\;
b_k = \sum_{j=0}^{M-1} 
A_{kj} \, x_j \, \mathsf{mod} \, 2
\;,
\end{equation} 
where the $M \times M$ binary matrix $A$ has the structure 
of a binary tree \cite{bravyi2002fermionic,seeley2012bravyi}.
Since it requires only $\mathcal{O}(\log_2 M)$-qubit operators, the BK transformations allows for more economical
encoding of fermionic operators onto qubit operators, with reduced cost for both measurements and quantum circuits.

\paragraph{Qubit reduction techniques.}

Lowering the number of qubits required to encode fermionic degrees of freedom, 
for example leveraging Hamiltonian symmetries \cite{bravyi2017tapering,setia2020reducing,faist2020continuous},
is an active and valuable research direction \cite{bravyi2017tapering,steudtner2018fermion}.
Since the Hilbert space of a single qubit is isomorphic to $\mathbb{C}^2$ and operators acting on different qubits 
commute with each other, it is natural to consider Abelian symmetry groups isomorphic to $\mathbb{Z}_2^{\times k}$.
Example of such symmetries are those generated by parities $(-1)^{N_\uparrow}$ and $(-1)^{N_\downarrow}$, 
proper rotations $C_2$, plane reflections $\sigma$ and inversion $i$.
In the tapering algorithm \cite{bravyi2017tapering}, such symmetries are detected leveraging the formalism of stabilizer groups \cite{gottesman1997stabilizer}:
the Hamiltonian is written as $\hat{H} = \sum_j c_j P_j$, where $P_j$ denotes an $M$-qubit Pauli operator, and symmetry groups $\mathcal{S}$ that
(i) are Abelian subgroups of the $M$-qubit Pauli group,
(ii) $-I \notin \mathcal{S}$,
and 
(iii) every element of $\mathcal{S}$ commutes with every Pauli operator $P_j$ in the qubit representation of $\hat{H}$
are considered.
While such a restriction limits the generality of the formalism, it provides an efficient algorithm, based on linear algebra on the $\mathbb{Z}_2$ field 
\cite{bravyi2017tapering}, to identify a set of generators $\tau_1 \dots \tau_k$ for $\mathcal{S}$, and a Clifford transformation $U$ such that 
$\tau_i = U^\dagger \, Z_i \, U$ for all $i=1 \dots k$.
An $M$-qubit wavefunction $\Psi$ is an eigenfunction of all symmetry operators with eigenvalues $s_i$ if $Z_i \, U | \Psi \rangle = s_i \, U | \Psi \rangle$, 
{i.e. $U | \Psi \rangle = | \Phi_{\bf{s}} \rangle \otimes | {\bf{s}} \rangle$. One can thus search for eigenfunctions of the projections 
$\hat{H}_{\bf{s}} = \mathrm{Tr}_{1 \dots k} \left[ \left( \mathbbm{1} \otimes |{\bf{s}}\rangle \langle {\bf{s}}| \right) U^\dagger H U \right]$ of $\hat{H}$ }
on the irreducible representation of $\mathcal{S}$ labeled by the eigenvalues ${\bf{s}}$.

The procedure is exemplified in Fig.~\ref{fig:representations}b.
At the end of the circuit, qubits 0,1 and 3 from above contain the irrep label and parities $p_1$, $p_3$ respectively, so the calculation of the energy of the H$_2$ molecule with the STO-6G basis using 4 qubits has resulted in the same energy but only 
using 1 qubit.

\subsubsection{Bosons in second quantization}
\label{sec:bosons_second}

In the previous section, we showed how to map spin-$1/2$ fermions to qubits. 
For many problems, it is necessary to simulate $d$-level particles with $d>2$, including 
bosonic elementary particles \cite{fisher1989boson}, 
spin-$s$ particles \cite{levitt2013spin}, 
vibrational modes \cite{wilson1980molecular} and electronic energy levels in molecules 
and quantum dots \cite{turro1991modern,hong2019overview}. 
Accordingly, several qubit-based quantum algorithms were recently developed for efficiently studying some of these systems, 
including nuclear degrees of freedom in molecules \cite{veis2016quantum,joshi2014estimating,teplukhin2019calculation,mcardle2019digital,sawaya2019quantum}, 
the Holstein model \cite{macridin2018electron,macridin2018digital} and quantum optics \cite{sabin2020digital,di2020variational}.

Mapping a $d$-level system to a set of qubits can be done in a variety of ways, 
and determining which encodings is optimal for a given problem has important practical implications.
The standard binary mapping refers to the familiar base-two numbering system, such that an integer $l=0\dots d-1$, corresponding to one of the $d$ levels of the system, 
is represented as $l = \sum_{i=0}^{n_q-1} x_i 2^i$, with $n_q = \lceil \log_2 d \rceil$, and mapped on the binary string ${\bf{x}}_l = (x_0 \dots x_{n_q-1})$. 
This simple and natural mapping has been used for qubit-based quantum simulation of truncated bosonic degrees of freedom \cite{veis2016quantum,joshi2014estimating,teplukhin2019calculation,mcardle2019digital,sawaya2019quantum,macridin2018electron,macridin2018digital,sabin2020digital,di2020variational}.
One mapping from classical information theory with particularly useful properties is called the
Gray or the reflected binary code. Its defining feature is that the binary strings ${\bf{x}}_{l+1}$, ${\bf{x}}_l$ encoding two consecutive integers $l+1$, $l$ 
have to differ by one digit only (i.e. the Hamming distance between the two bitstrings is $1$). This encoding is especially favorable for tridiagonal operators 
with zero diagonals, and requires $n_q = \lceil \log_2 d \rceil$ qubits.
There exist encodings that make less efficient use of quantum resources, requiring more than $\lceil \log_2 d \rceil$ qubits, but usually allow for fewer quantum operations. Among them one finds the unary encoding, using $2^d$ qubits and mapping integers $l$ onto binary strings 
$\left( {\bf{x}}_l \right)_m = \delta_{lm}$. Previous proposals for digital quantum simulation of bosonic degrees of freedom have used the unary encoding.
The standard binary, Gray and unary codes are described in detail and compared in the literature \cite{sawaya2020resource,sawaya2020near}.
The most efficient encoding choice is often significantly dependent on the application at hand, due to the complicated and delicate interplay between 
Hamming distances, sparsity patterns, bosonic truncation, and other properties of the Hamiltonian and other operators, and 
it is sensitive to the number $d$ of levels of the system
\cite{veis2016quantum,joshi2014estimating,teplukhin2019calculation,mcardle2019digital,sawaya2019quantum,macridin2018electron,macridin2018digital,sabin2020digital,di2020variational}. 

\subsubsection{Alternative approaches}
\label{sec:fermions_bosons_alternatives}

For local Hamiltonians in one spatial dimension, the JW transformation allows mapping of a local theory of fermions onto a local theory of spins. 
In higher dimensions, however, the JW transformation gives rise to non-local coupling between spins.
The possibility of reducing this non-local coupling has been explored by many authors, including Bravyi and Kitaev \cite{bravyi2002fermionic}.
Similar ideas were explored by Ball \cite{ball2004fermions} and Cirac and Verstraete \cite{verstraete2005mapping}.
The latter, in particular, achieved a mapping between local Fermi and local qubit Hamiltonians by introducing extra degrees of freedom 
in the form of Majorana fermions, which interact locally with the original ones.
Other approaches have included the use of Fenwick trees \cite{havlivcek2017operator} and graph theoretical tools \cite{setia2019superfast}.

Generalizations of the tapering qubit reduction formalism to more general discrete and continuous symmetries is known \cite{setia2020reducing,faist2020continuous}, 
and symmetries have been recognised as an important ingredient in the implementation of error correction algorithms \cite{bonet2018low,mcardle2019error}.
Recently, techniques reducing the number of qubits by half focusing on the seniority-zero sector of the Hilbert space and partitioning an electronic
system into classically correlated spin-up and spin-down sectors have also been proposed \cite{elfving2021simulating,eddins2021doubling}. 
As illustrated by the many different techniques 
referred to above, this continues to remain an important research direction for a variety of reasons: first, qubit
reduction techniques allow for the simulation of  problems with less
quantum resources and thus to make use of near-term hardware;
furthermore, they help enforce exact properties of electronic wavefunctions (e.g. particle number and spin conservation
and, in presence of molecular point-group symmetries, 
labeling ground and excited states by irreducible representations).

\subsection{Simulation of Hamiltonian dynamics}
\label{sec:dynamics}

As discussed in Section \ref{sec:complexity}, the simulation of Hamiltonian dynamics, 
i.e. the solution of the time-dependent Schr\"{o}dinger equation,
is a BQP-complete problem, and thus a natural application for a quantum computer.

In this Section, we present some important quantum algorithms for the simulation of Hamiltonian dynamics.
Given a Hamiltonian $\hat{H}$ acting on a set of $n$ qubits, we say \cite{childs2004quantum} 
that $\hat{H}$ can be {\bf{efficiently simulated}} (to an accuracy $\varepsilon$)    if one can produce a quantum circuit $\hat{U}$ such that
\begin{equation}
\label{eq:hamiltonian_simulation}
\| \hat{U} - e^{-it \hat{H}} \| < \varepsilon
\quad,
\end{equation}
and $\hat{U}$ comprises a number of gates scaling at most polynomially with $n$, $t$ and $\varepsilon^{-1}$.
In the remainder of this Section, we will describe quantum algorithms to simulate the electronic structure Hamiltonian, 
in the sense of Eq.~\eqref{eq:hamiltonian_simulation}.
These algorithms will be classified, according to the underlying mathematical formalism, into 
product formula \cite{lloyd1996universal,abrams1997simulation}, quantum walk \cite{childs2009universal,berry2015hamiltonian} 
and linear combination of unitary operators (LCU) algorithms \cite{low2017optimal,berry2015simulating,low2019hamiltonian,childs2018toward}.

\subsubsection{Product formula algorithms}

\begin{figure}
\includegraphics[width=0.9\textwidth]{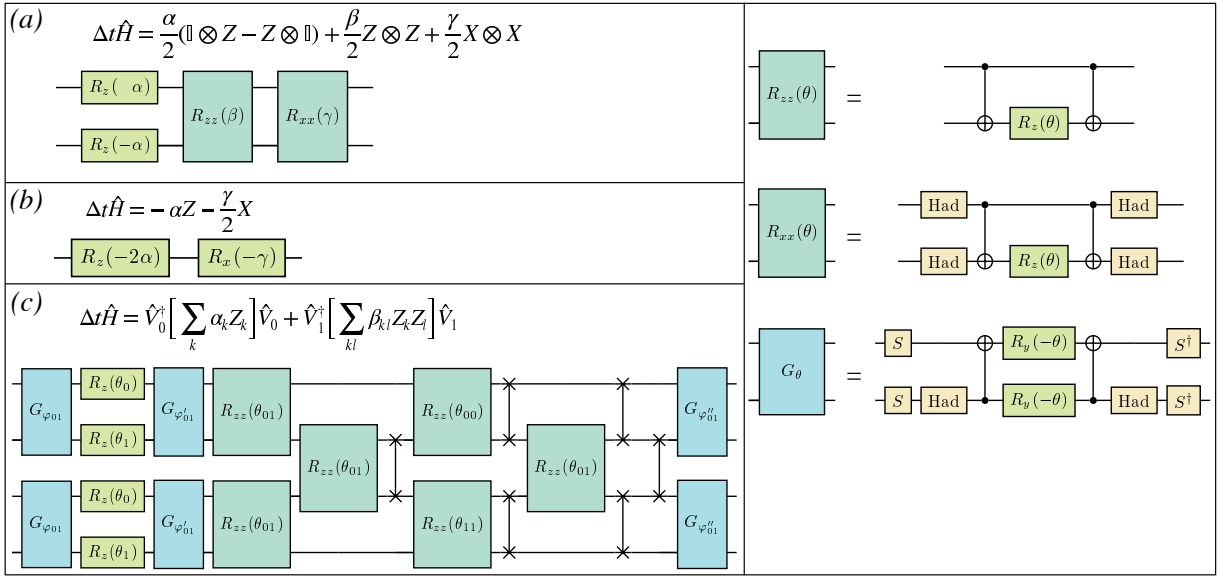}
\caption{
Left:
$(a)$ quantum circuit to simulate a Trotter step of time evolution under the Hamiltonian of $\ce{H_2}$ at STO-6G level in parity representation with two-qubit reduction
$(b)$ same as $(a)$ but accounting for $D_{\infty h}$ symmetry via reduction of a third qubit.
$(c)$ quantum circuit to simulate a Trotter step of time evolution under the Born-Oppenheimer Hamiltonian in a set of 2 spatial orbitals and Jordan-Wigner representation, 
using a low-rank decomposition with $N_\gamma = 1$ term. Blue, teal and green blocks indicate Givens rotations, 
exponentials of $Z \otimes Z$ and $X \otimes X$ Pauli operators, and single-qubit $Z$ rotations respectively. Angles $\varphi_{kl}$ parametrize Givens rotations, and angles $\theta_k$, $\theta_{kl}$ are functions of the Hamiltonian coefficients $\alpha_k$, $\beta_{kl}$ respectively. 
Right:
Implementation of the exponentials $R_{zz}(\theta)$ (top),
$R_{xx}(\theta)$ (middle), and of a Givens rotation (bottom).
}
\label{fig:trotter}
\end{figure}

Given a Hamiltonian operator $\hat{H} = \sum_{\ell=1}^L \hat{h}_\ell$, product formulas produce an approximation to 
$e^{-i t \hat{H}}$ using a product of exponential operators $e^{-i t \hat{h}_\ell}$. 
This is achieved by dividing the time interval $[0,t]$ into a large number $n_T$ of steps,
\begin{equation}
\hat{U}(t) = e^{-it \hat{H}} = \prod_{i=0}^{n_T-1} \hat{U}(\Delta t)
\quad,\quad
\Delta t = \frac{t}{n_T}
\quad,
\end{equation}
and approximating each of the operators $\hat{U}(\Delta t)$.
An important example is the primitive approximation \cite{trotter1959product,suzuki1976generalized}
\begin{equation}
\label{eq:primitive_trotter}
\hat{U}(\Delta t) = \prod_{\ell} e^{-i \Delta t \hat{h}_\ell} + \mathcal{O}(\Delta t^2) 
\quad,\quad
\hat{U}_p(\Delta t) 
\equiv
\prod_{\ell} e^{-i \Delta t \hat{h}_\ell}
\quad.
\end{equation}
The accuracy of the primitive approximation Eq.~\eqref{eq:primitive_trotter} can be estimated observing that 
\begin{equation}
\| \hat{U}(t) - \hat{U}_p(\Delta t)^{n_T} \| 
= \gamma_p \, n_T \, \Delta t^2 + \mathcal{O}(t^3/n_T^2)
\quad,\quad
\gamma_p = \frac{1}{2} \, \sum_{\ell< \ell^\prime} \| [ \hat{h}_\ell , \hat{h}_{\ell^\prime} ] \|
\quad .
\end{equation}
Therefore, accuracy $\varepsilon$ is attained for $n_T = \mathcal{O}( \gamma_p \, t^2 / \varepsilon )$.
More accurate approximations can be achieved relying on the Trotter-Suzuki formulas 
\cite{trotter1959product,suzuki1976generalized,suzuki1991general,childs2019theory}, 
which are defined recursively as
\begin{equation}
\label{eq:LTS_2}
\begin{split}
\hat{U}_{2k+2}(\Delta t) 
&= 
\hat{U}_{2k}^2 \left( a_{2k} \Delta t \right)
\hat{U}_{2k} \left( (1-4 a_{2k}) \Delta t \right)
\hat{U}_{2k}^2 \left(a_{2k} \Delta t \right) \quad, \\
\hat{U}_{2}(\Delta t) 
&= 
\left[ \; \prod_{\ell=L}^1 e^{-i \frac{\Delta t}{2} \hat{h}_\ell} \; \right] 
\left[ \; \prod_{\ell=1}^L e^{-i \frac{\Delta t}{2} \hat{h}_\ell} \; \right] \quad,\quad
a_{2k} = \frac{1}{4-4^{\frac{1}{2k+1}}} \quad. \\
\end{split}
\end{equation}
Since $\hat{U}_{2k}(\Delta t) = e^{ - i \Delta t \hat{H} }  + \mathcal{O}(\Delta t^{2k+1})$, 
accuracy $\varepsilon$ is achieved for $n_T = t^{ 1 + \frac{1}{2k}} \varepsilon^{- \frac{1}{2k}}$.
As seen, the cost of simulating Hamiltonian dynamics with product formulas has power-law scaling in $t$ and $\varepsilon^{-1}$.
To achieve efficient simulation, however, the number $L$ of Hamiltonian terms and the cost of simulating $e^{-i \Delta t \hat{h}_\ell}$ 
has to scale as $\mbox{poly}(n)$.
This can be achieved, for example, mapping the electronic structure Hamiltonian Eq.~\eqref{eq:hamiltonian_born_oppenheimer} 
onto an $M$-qubit operator $\hat{H} = \sum_{\ell=1}^L c_\ell \hat{P}_\ell$, as outlined in Section \ref{sec:mapping_to_qubits},
and exponentiating individual $M$-qubit Pauli operators $\hat{P}_\ell$ with the circuit presented in Section~\ref{sec:hardware},
and exemplified in Fig.~\ref{fig:trotter}.
The resulting scaling is between $\mathcal{\tilde{O}}(M^4)$ and $\mathcal{O}(M^5)$ per step \cite{childs2018toward,motta2018low}.

Alternative approaches exist, which can be used to lower the computational cost.
In particular, Aharonov and Ta-Shma \cite{aharonov2003adiabatic,berry2007efficient}
introduced an algorithm for the quantum simulation of $d$-sparse Hamiltonians,
based on a combination of graph coloring and Trotter decomposition.
More recently, various authors \cite{poulin2014trotter,babbush2017low,jiang2018quantum,kivlichan2018quantum,motta2018low,google2020hartree,matsuzawa2020jastrow} 
proposed product formulas based on low-rank representations of the electron-repulsion integral,
$(pr|qs) = \sum_{\gamma=1}^{N_\gamma} L^\gamma_{pr} L^\gamma_{qs}$. Such representations
can be efficiently computed from a density fitting approximation \cite{
Whitten:1973:4496,
Dunlap:1977:81,
Dunlap:1979:3396,
Feyereisen:1993:359,
Komornicki:1993:1398,
Vahtras:1993:514,
Rendell:1994:400,
Kendall:1997:158,
Weigend:2002:4285}, a Cholesky decomposition
\cite{
Beebe:1977:683,
Roeggen:1986:154,
Koch:2003:9481,
Aquilante:2007:194106,
Aquilante:2009:154107,
motta2019efficient,
peng2017highly},
or a multi-unitary tensor hypercontraction representation 
\cite{hohenstein2012tensor,parrish2012tensor,parrish2013exact,matsuzawa2020jastrow,cohn2021quantum}
and typically feature $N_\gamma = \mathcal{O}(M)$ terms.
In this framework, the electronic structure Hamiltonian is represented as
\begin{equation}
\hat{H} = E_0 + \hat{V}_0^\dagger \left[ \sum_p t_p \hat{n}_p \right] \hat{V}_0
+ \sum_\gamma \hat{V}_\gamma^\dagger 
\left[ \sum_{pq} v^\gamma_{pq} \hat{n}_p \hat{n}_q \right] 
\hat{V}_\gamma
\quad,\quad
\hat{V}_\gamma = e^{ \sum_{pq} A^\gamma_{pq} \crt{p} \dst{q}. }
\quad,
\end{equation}
and the time evolution operators $e^{-i \Delta t \hat{H}}$ as 
\begin{equation}
e^{-i \Delta t \hat{H} } \simeq 
\prod_\gamma
\left[ 
\hat{V}_\gamma^\dagger
\,
\Bigg( \prod_{pq} e^{-i \Delta t \, v^\gamma_{pq} \, \hat{n}_p \hat{n}_q } \Bigg)
\hat{V}_\gamma
\right] 
\,
\prod_p e^{-i \Delta t \, \hat{n}_p }
\quad.
\end{equation}
Exponentials of one-body operators such as $\hat{V}_\gamma$ can be decomposed 
into a product of up to $\mathcal{O}(M^2)$ Givens transformations \cite{jiang2018quantum,kivlichan2018quantum}
with standard linear algebra operations.
Each of these Givens rotations can be implemented, in the JW representation,
with a two-qubit gate acting on a pair of adjacent qubits. 
Exponentials of number and products of number operators can be implemented \cite{jiang2018quantum}
with $R_z(\varphi) = \exp(-i \varphi/2 Z)$ and $R_{zz}(\varphi) = \exp(-i \varphi/2 \, Z \otimes Z)$ gates, as exemplified in Fig.~\ref{fig:trotter}. 
The resulting scaling is $\mathcal{O}(N_\gamma M^2) = \mathcal{O}(M^3)$ \cite{kivlichan2018quantum,motta2018low}.
Other approaches to implement time evolution with product formulas have explored
alternatives to the high-order formulas Eq.~\eqref{eq:LTS_2} having a more favorable scaling with $k$ \cite{low2019well},
as well as grouping commuting Pauli terms \cite{poulin2014trotter} and divide-and-conquer strategies \cite{haah2018quantum}.
Recently, Childs and Su \cite{childs2019nearly} revisited approaches based on product formulas, 
and demonstrated that their performance is more favorable than  suggested by simple error bounds. 
Techniques of circuit recompilation aimed at generating optimal circuit are also intensely investigated
\cite{whitfield2011simulation,Hastings15,childs2018faster,campbell2018random}.
The design and improvement of product formulas, and the systematic assessment of their accuracy and computational 
cost for molecular problems, is thus an active and valuable research area at the interface between quantum computation 
and chemistry \cite{poulin2014trotter,cao2019quantum,bauer2020quantum}.

\subsubsection{Quantum walks}
\label{sec:qw}

Hamiltonian simulation by product formulas has cost scaling {\bf{super-linearly}} with $t$. Such an observation leads to question of whether a better scaling can be achieved.
For a general Hamiltonian, a scaling sub-linear in $t$ cannot be achieved, due to the no-fast-forwarding theorem \cite{berry2007efficient,childs2009limitations}.
On the other hand, a number of algorithms achieving linear scaling with $t$ are known. 
The earliest example is represented by quantum walk algorithms \cite{kempe2003quantum,venegas2012quantum,berry2009black,childs2010simulating,childs2010relationship}.

In the context of Hamiltonian simulation by quantum walks, we consider a Hamiltonian operator
$\hat{H} = \sum_\lambda \lambda \ket{\lambda} \bra{\lambda}$ acting on a Hilbert space $\mathcal{H}$ and having eigenvalues $\lambda \in [-1,1]$.
Such a condition can always be satisfied by computing an upper bound $\zeta \geq \| \hat{H} \|$ for the norm of $\hat{H}$, 
and rescaling the Hamiltonian and simulation time accordingly.
Simulating Hamiltonian dynamics 
by a quantum walk consists in introducing a unitary operator $\hat{W}$, acting on an extended Hilbert space $\mathcal{H}_e \supset \mathcal{H}$,
whose spectrum is connected with that of $e^{-i \Delta t \hat{H}}$ by an invertible transformation.
Such a unitary transformation has the form \cite{berry2009black,childs2010simulating,childs2010relationship}
\begin{equation}
\label{eq:qw}
\hat{W} = i \hat{S} \left( 2 \hat{T} \hat{T}^\dagger - \mathbbm{1} \right)
\quad,
\end{equation}
where $\hat{T} : \mathcal{H} \to \mathcal{H}_e$ is an isometry, $\hat{S} : \mathcal{H}_e \to \mathcal{H}_e$ a unitary operator such that
$\hat{S}^2 = \mathbbm{1}$, and
\begin{equation}
\label{eq:q_walk_0}
\hat{T}^\dagger \hat{S} \hat{T} = \hat{H} 
\quad.
\end{equation}
Thanks to these properties, it can be proved \cite{berry2009black,childs2010relationship} that $\hat{W}$ 
leaves a set of 2-dimensional subspaces 
\begin{equation}
\mathcal{S}_\lambda = \mathrm{span} \left\{ \hat{T} | \lambda \rangle , \hat{S} \hat{T} | \lambda \rangle \right\}
\end{equation}
invariant, and has eigenvalues and eigenvectors
\begin{equation}
\hat{W} \ket{ \mu_\pm(\lambda) } = \mu_\pm(\lambda) \ket{ \mu_\pm(\lambda) }
\quad,\quad
\mu_\pm(\lambda) = \pm e^{ \pm i \arcsin(\lambda)}
\quad,\quad
| \mu_\pm(\lambda) \rangle = \frac{\mathbbm{1} + i \mu_\pm(\lambda) \hat{S}}{\sqrt{2 (1-\lambda^2)}} \, \hat{T} | \lambda \rangle
\quad .
\end{equation}
The operators $\hat{S}$ and $\hat{T}$ are constructed assuming the existence of an oracle, 
i.e. a unitary operator giving access to the binary representation of the matrix elements of $\hat{H}$ in a suitable basis
\cite{berry2009black,childs2010relationship}, and the spectrum of $\hat{W}$ is converted into that of $e^{-i t \hat{H}}$ 
using a subroutine based on the quantum phase estimation algorithm (see Section \ref{sec:applications_of_time_evolution}).
The quantum walk approach, though achieving linear scaling with $t$, retains algebraic scaling with $\varepsilon$,
which was improved by LCU-based algorithms.

\subsubsection{The linear combination of unitary operators: LCU lemma}
\label{sec:lcu}

Product formulas and quantum walks permit to 
approximate the time evolution operator by a sequence of unitary operations.
A breakthrough in quantum simulation came from the realization
that more accurate approximations can be achieved using {\bf{non-unitary approximations}} \cite{berry2014exponential,berry2015simulating,low2016methodology,low2017optimal,low2019hamiltonian}.
In particular, such non-unitary approximations often take the
form of linear combinations of unitary operations.

\begin{figure}[h!]
\includegraphics[width=0.9\textwidth]{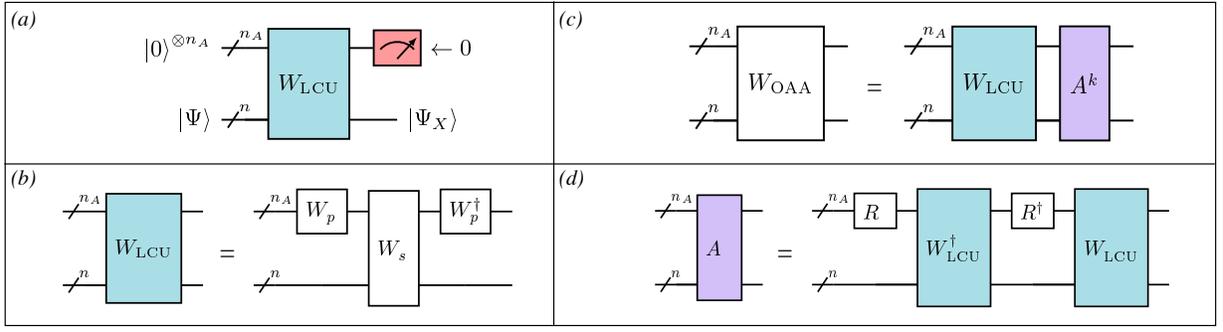}
\caption{Schematic representation of the algorithm for the probabilistic implementation of an LCU (a), structure of the LCU unitary (blue block) as composition of a preparation and a selection unitary (b), 
schematic representation of the OAA algorithm (c), 
and implementation of the OAA unitary (purple block) as a composition of the LCU unitary and a reflector unitary transformation (d). Slashes denote multi-qubit registers.}
\label{fig:lcu_general}
\end{figure}

In order to describe several important quantum algorithms 
based on non-unitary approximations of the time evolution operator, in this section we present a theoretical result 
known as LCU lemma \cite{berry2015simulating}, showing how
to apply linear combination of $L$ unitary operators
to a set of qubits prepared in a state $|\Psi \rangle$, 
\begin{equation}
\label{eq:lcu_target}
| \Psi \rangle \mapsto \frac{ \hat{X} | \Psi \rangle}{\| \hat{X} \Psi \|} \equiv | \Psi_X \rangle
\quad,\quad
\hat{X} = \sum_{\ell=0}^{L-1} \alpha_\ell \hat{U}_\ell
\quad.
\end{equation}
Since $\hat{X}$ is not a unitary operator, it is not straightforward to represent the map \eqref{eq:lcu_target} as a quantum circuit. 
On the other hand, such a representation is highly desirable, as it extends the reach of quantum computation to non-unitary operators.
The LCU lemma provides a strategy for implementing the transformation \eqref{eq:lcu_target} with probability $p$, 
based on the quantum circuit in Figure \ref{fig:lcu_general}a.
A register of $n_A = \lceil \log_2(L) \rceil$ ancillary qubits is prepared in $|0 \rangle^{\otimes n_A}$ and coupled to a register of $n$ qubits, 
prepared in the state $| \Psi \rangle$.
The ancillae are manipulated with a preparation unitary
\begin{equation}
\label{eq:lcu_p}
\hat{W}_p | 0 \rangle 
= 
\sum_{\ell=0}^{L-1} \sqrt{\frac{\alpha_\ell}{\alpha}} \, | \ell \rangle
\quad,\quad
\alpha = \sum_{\ell=0}^{L-1} \alpha_\ell
\quad,
\end{equation}
and coupled with the qubits of the main register by a selection unitary
\begin{equation}
\hat{W}_s 
= 
\sum_{\ell=0}^{L-1} {\hat{U}_\ell \otimes | \ell \rangle \langle \ell |}
\quad.
\end{equation}
The transformation $\hat{W}_p$ is subsequently reversed, and the ancillae are measured. If the outcome of the measurement is $(0 \dots 0)$,
which happens with probability $p = \| \hat{X} \psi \|^2 / \alpha^2$, the qubits of the main register collapse onto the state $| \Psi_X \rangle$.
The number of ancillae, $n_A = \lceil \log_2(L) \rceil$ scales logarithmically with $n$ provided that
$L = \mathrm{poly}(n)$. The unitary $\hat{W}_p$ can require up to $2^{n_A} = \mathrm{poly}(n)$ gates \cite{barenco1995elementary}. 
Similarly, if every unitary $\hat{U}_\ell$ can be controlled at cost $c_\ell = \mathrm{poly}(n)$, 
then $\hat{W}_s$ can be implemented at cost $\mathrm{poly}(n)$ \cite{barenco1995elementary}.
The main limitation of the LCU algorithm is the success probability, which decays as $1-(1-p)^k = 2^{- \mathcal{O}(k)}$ when the circuit in
Fig.~\ref{fig:lcu_general}a is applied $k$ times consecutively. 
{When $\hat{X}$ is a unitary operator \cite{berry2015hamiltonian}, as in Hamiltonian evolution, the} success probability $p$ can be increased using the procedure called oblivious amplitude amplification (OAA) 
\cite{brassard1997exact,grover1998quantum,berry2014exponential}.
The OAA is described by the quantum circuit in Figure \ref{fig:lcu_general}, where $\hat{W}_{\mathrm{LCU}} = \hat{W}_p^\dagger \hat{W}_s \hat{W}_p$ is the LCU unitary, 
and {$\hat{R} = \mathbbm{1} \otimes \left( \mathbbm{1} - 2 | 0 \rangle \langle 0 | \right)$ }reflects ancillae around the state $| 0 \rangle \langle 0 |$.
The circuit $\hat{A}^k \hat{W}_{\mathrm{LCU}}$ leads to a state of the form
\begin{equation}
\hat{U} | 0 \rangle | \Psi \rangle = \sqrt{p_k} \, {\frac{\hat{X} | \Psi \rangle}{\| \hat{X} \Psi \|} \otimes | 0 \rangle^{\otimes n_A}} + \sqrt{1-p_k} \, | \Phi^\perp \rangle
\quad,\quad
p_k = \sin^2 \big( (2k+1) \theta \big)
\quad,\quad
{\theta = \arcsin \sqrt{p}}
\quad,
\end{equation}
and $p_k \geq 1-\delta$ provided that $k \geq \mathcal{O}( 1 / \sqrt{p} )$.
We remark that, when $\hat{X}$ is not a unitary operator, OAA gives Chebyshev polynomials of 
that operator \cite{berry2015hamiltonian}.

\subsubsection{LCU based algorithms}
\label{sec:taylor}

\begin{figure}[h!]
\includegraphics[width=0.6\textwidth]{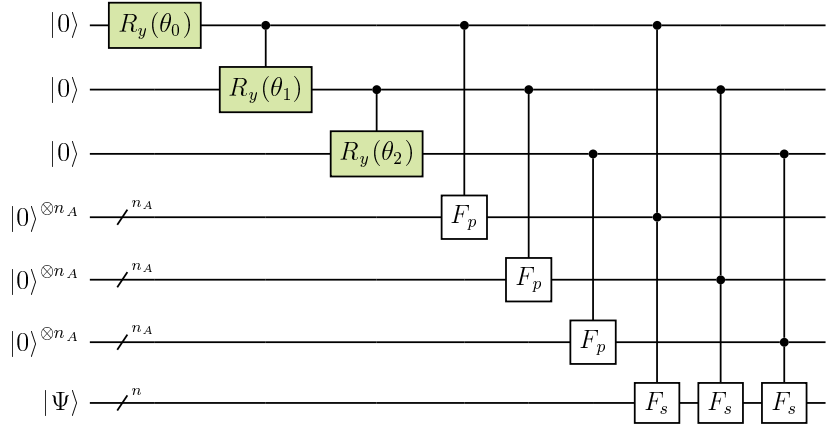}
\caption{
{Preparation ($y$ rotations, multi-qubit gates $\mathsf{c}\hat{F}_p$) and selection (multi-qubit $\mathsf{c} \hat{F}_s$ gates) unitaries for Hamiltonian simulation based on a truncated Taylor series. Here $K=3$ and $n_A = \log_2(L)$. 
The angles in the $y$ rotations are chosen to reproduce the state
$\sum_m \sqrt{\Delta t^m/m!} { \, | 0 \rangle^{K-m} | 1 \rangle^{m}}$. The operation $\mathsf{c}\hat{F}_p$ is defined as $\mathsf{c}\hat{F}_p | 0 \rangle | k \rangle = V^k | 0 \rangle | k \rangle$, with $V | 0 \rangle = \sum_\ell \sqrt{\alpha_\ell} | \ell \rangle$, and the 
operation $\mathsf{c} \hat{F}_s$ as $\mathsf{c} \hat{F}_s | \psi \rangle | \ell \rangle | k \rangle = (-i \hat{U}_\ell)^k | \psi \rangle | \ell \rangle | k \rangle$.}}
\label{fig:taylor_circuit}
\end{figure}

In 2014, Berry et al \cite{berry2015simulating} introduced a method for Hamiltonian simulation, based on the Taylor series representation of the time evolution operator,
to achieve a computational cost scaling linearly in $t$ and logarithmically in $\varepsilon^{-1}$ (i.e. exponentially better than product formulas and quantum walks).
The algorithm focuses on Hamiltonian operators that can be written as LCUs, $\hat{H} = \sum_{\ell=0}^{L-1} \alpha_\ell \hat{U}_\ell$. {For simplicity, we also assume that
$C = \sum_\ell \alpha_\ell = 1$, which is equivalent to rescaling energy and time units, $\hat{H} \to C^{-1} \hat{H}$ and $t \to C t$.}
The time interval $[0,t]$ is divided in $r$ steps of duration $\Delta t = t / r$, and the operator $e^{-i \Delta t \hat{H}}$ is expanded in Taylor series to order $K$,
\begin{equation}
\label{eq:lcu_taylor}
e^{-i \Delta t \hat{H}} 
\simeq 
\hat{V}_K(\Delta t)
= 
\sum_{m=0}^{K} \frac{(-i \Delta t)^m}{m!} \hat{H}^m
= 
\sum_{m=0}^{K} \sum_{\ell_0 \dots \ell_{m}=0}^{L-1} 
\frac{(-i \Delta t)^m}{m!} \alpha_{\ell_0} \dots \alpha_{\ell_{m}} \hat{U}_{\ell_0} \dots \hat{U}_{\ell_{m}}
\quad.
\end{equation}
Eq.~\eqref{eq:lcu_taylor} is an LCU representation of $e^{-i \Delta t \hat{H}}$, which can be probabilistically applied 
relying on a particular implementation \cite{berry2015simulating} of the LCU lemma presented in Section \ref{sec:lcu}, where $K (1+\log_2 L)$ ancillae are used,
as shown in Fig.~\ref{fig:taylor_circuit}.
The first $K$ ancillae are prepared in the state $\sum_m \sqrt{\Delta t^m/m!} { \, | 0 \rangle^{K-m} | 1 \rangle^{m}}$ using $\mathcal{O}(K)$ controlled single-qubit rotations,
and the remaining $\log_2(L)$ groups of $K$ ancillae are prepared in the normalized state $\sum_\ell \sqrt{\alpha_\ell} | \ell \rangle$ using $\mathcal{O}(KL)$ gates.
The other basic component is the selection unitary, which maps states of the form {$\ket{\psi} \ket{\ell_k} \dots \ket{\ell_1} \ket{k}$} to 
{$(-i)^k \hat{U}_{\ell_1} \dots \hat{U}_{\ell_k} \ket{\psi} \ket{\ell_k} \dots \ket{\ell_1} \ket{k}$} at the cost of $\mathcal{O}( L ( n + \log_2L) K)$ operations \cite{berry2015simulating}.
An important aspect of this algorithm is the order $K$ of the polynomial approximating $e^{-i \Delta t \hat{H}}$ with accuracy $\varepsilon/K$, which scales as \cite{berry2015simulating}
\begin{equation}
K = \mathcal{O}\left( \frac{ \log \left( \frac{t}{\varepsilon} \right) }{ \log \log \left( \frac{t}{\varepsilon} \right) } \right) \quad.
\end{equation}
The logarithmic dependence of the computational cost on $\varepsilon^{-1}$ thus arises from the rapid convergence of the Taylor series,
as well as on the specific use of ancillary qubits and controlled operations described in the previous paragraph.

\subsubsection{Quantum signal processing and qubitization}

The qubitization algorithm, developed by Low and Chuang \cite{low2019hamiltonian}, uses the LCU decomposition of the Hamiltonian 
and the quantum signal processing (QSP) technique \cite{low2016methodology,low2017hamiltonian} to achieve a computational cost where the dependence on $t$ and $\log \varepsilon^{-1}$
is \important{additive rather than multiplicative}, $\mathcal{O}(t+\log \varepsilon^{-1})$, which is optimal in both accuracy and time \cite{low2017hamiltonian}.

\begin{figure}[b!]
\includegraphics[width=\textwidth]{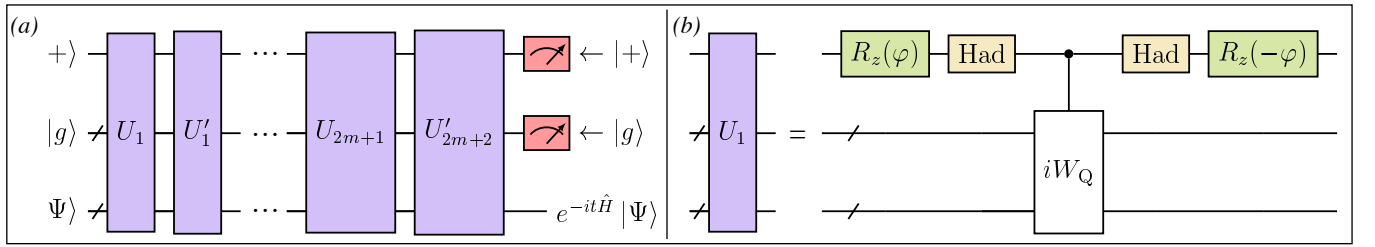}
\caption{Quantum circuit for single-ancilla quantum signal processing, where $U_k = U_{\varphi_k}$ and $U^\prime_{k} = U^\dagger_{\varphi_k+\pi}$ for brevity (a) and structure of the gate $\hat{U}_\varphi$ in terms of single-qubit operations and the qubiterate $\hat{W}_{\mathrm{Q}}$ (b).}
\label{fig:qsp}
\end{figure}

Given a Hamiltonian operator $\hat{H}$ acting on a Hilbert space $\mathcal{H}$ and having norm $\| \hat{H} \| \leq 1$,
the starting point of qubitization is the construction of a unitary operator $\hat{W}_{\mathrm{Q}}$, called qubiterate. 
The qubiterate acts on an extended Hilbert space $\mathcal{H}^\prime \otimes \mathcal{H}$ and is an encoding for $\hat{H}$, in the sense that
\begin{equation}
\label{eq:qubitization_0}
{
\langle \phi | \langle g | \hat{U} | \psi \rangle | g \rangle = \langle \phi | \hat{H} | \psi \rangle
}
\quad
\mbox{for some $|g \rangle \in \mathcal{H}^\prime$ and any $\ket{\phi}, \ket{\psi} \in \mathcal{H}$}
\quad.
\end{equation}
Moreover, for any eigenvector of $\hat{H}$, $\hat{H} \ket{\lambda} = \lambda \ket{\lambda}$, one has 
\begin{equation}
\label{eq:qubitization_1}
\hat{W}_{\mathrm{Q}} | g_\lambda \rangle = \lambda | g_\lambda \rangle - \sqrt{1-\lambda^2} \ket{ g_\lambda^\perp }
\quad,\quad
\hat{W}_{\mathrm{Q}} | g_\lambda^\perp \rangle = \lambda | g^\perp_\lambda \rangle + \sqrt{1-\lambda^2} | g_\lambda \rangle
\quad,
\end{equation}
where {$| g_\lambda \rangle = \ket{\lambda} \ket{g}$} and $\ket{ g_\lambda^\perp }$ is defined by the first of Eq.~\eqref{eq:qubitization_1}.
In other words, the qubiterate leaves the two-dimensional subspaces $\mathcal{S}_\lambda$ spanned by $| g_\lambda \rangle$ and $| g_\lambda^\perp \rangle$ invariant, 
so that its restriction to $\oplus_\lambda \mathcal{S}_\lambda$ is a direct sum of $Y$ rotations acting on the subspaces individually $\mathcal{S}_\lambda$, 
\begin{equation}
\hat{W}_{\mathrm{Q}} = \bigoplus_\lambda e^{-i \theta_\lambda Y_\lambda} \quad,\quad \theta_\lambda = \mbox{arccos}(\lambda)
\quad.
\end{equation}
The qubiterate is used to construct an operator $\hat{V}_{\vec{\varphi}}$, which in turn is used to approximate $e^{-it \hat{H}}$,
by a QSP \cite{low2016methodology,low2017hamiltonian}. The single-ancilla QSP is described by the quantum circuit in Fig.~\ref{fig:qsp}.

The operator $\hat{V}_{\vec{\varphi}}$ has the form
\begin{equation}
\hat{V}_{\vec{\varphi}} = \prod_{m=0}^{K} \hat{U}^\dagger_{\varphi_{2m+2}+\pi} \hat{U}_{\varphi_{2m+1}}
\quad,\quad
\hat{U}_\phi 
{= \left[ \mathbbm{1} \otimes R_z(-\varphi) \, \mathrm{Had} \right] \mathsf{c} \big(iW_{\mathrm{Q}} \big) \left[ \mathbbm{1} \otimes \mathrm{Had} \, R_z(\varphi) \right]}
\quad,
\end{equation}
{where $\mathsf{c} \big(iW_{\mathrm{Q}} \big)$ is the controlled version of the 
gate $iW_{\mathrm{Q}}$.} It can be shown that \cite{low2019hamiltonian}
\begin{equation}
\label{eq:qubitization_5}
\hat{V}_{ \vec{\varphi} } 
=
\sum_{\lambda,\pm}
{| \lambda_\pm \rangle \langle \lambda_\pm |
\otimes
u( \vec{\varphi} , \theta_{\lambda,\pm} )}
\quad,
\end{equation}
where $| \lambda_\pm \rangle$ is an eigenvector of $\hat{W}_{\mathrm{Q}}$ with eigenvalue $e^{ \pm i \theta_\lambda}$, and $u$ a single-qubit operator.
Provided that the number of angles $\vec{\varphi}$ is sufficiently large, $K = \mathcal{O}(t + \log \varepsilon^{-1})$, these angles can be 
set to values that are efficiently computable on a classical computer \cite{low2019hamiltonian,dong2021efficient,martyn2021grand,chao2020finding} so that
\begin{equation}
{\langle \chi | \langle g | \langle + | \hat{V}_{ \vec{\phi}} | \psi \rangle | g \rangle | + \rangle \simeq \langle \chi | e^{-it \hat{H}} | \psi \rangle}
\end{equation}
with error bounded by $\varepsilon$.
Qubitization is a systematic framework, that offers a concrete procedure to simulate Hamiltonian dynamics with optimal complexity with respect to $t$ and $\varepsilon$.
In recent years, its complexity with respect to the number of spin-orbitals was improved using low-rank \cite{berry2019qubitization} and tensor hypercontraction \cite{lee2020even} techniques.
Product formulas and Taylor series techniques allow simulation of
Schr\"{o}dinger equations with a time-dependent Hamiltonian $\hat{H}(t)$.
On the other hand, the qubitization technique is formulated for time-independent Hamiltonians, and its extension to time-dependent Hamiltonians is an open and challenging research problem.
An important consequence of this difference is that product formulas and Taylor series techniques allow simulation in
the interaction picture of quantum mechanics. This is especially desirable for electronic structure problems, where the Born-Oppenheimer Hamiltonian can be written as the sum $\hat{H} = \hat{T} + \hat{V}$ of a one-body operator and of a two-body operator, so that the unitary transformation 
\begin{equation}
\Psi(t) \to \Psi_{I}(t) = e^{ \frac{t}{i\hbar} \hat{T}} \Psi_t
\end{equation}
leads to a Schr\"{o}dinger equation with a time-dependent Hamiltonian $\hat{V}_I(t) = e^{ \frac{t}{i\hbar} \hat{T}} \hat{V} e^{ - \frac{t}{i\hbar} \hat{T}}$. It is important
to remark that, since $\hat{T}$ is a one-body operator, 
the interaction term $\hat{V}$, and any other molecular property described by a $k$-body operator, can be exactly 
and efficiently transformed to the interaction picture, $\hat{V} \to \hat{V}_I(t)$.

\subsubsection{Applications of Hamiltonian dynamics, and open problems}
\label{sec:applications_of_time_evolution}

In the previous Section, we examined some {emerging} quantum algorithms for Hamiltonian dynamics simulation. Hamiltonian dynamics simulation
is one of the most compelling applications for a quantum computer, as it lies in the complexity class BQP.
The study, implementation and application of algorithms for Hamiltonian dynamics simulation is thus a relevant research area, 
and one that has significant potential to yield the first
relevant quantum simulations of chemical systems \cite{childs2018toward}.
Nevertheless, a careful consideration of the actual computational cost and of the accuracy of such algorithms,
with particular attention to chemical problems, is a necessary requirement to understand, design and carry out such simulations.

While product formula algorithms have the highest asymptotic computational cost, compared against quantum walks and LCU (linear combination of unitary operators) based algorithms,
the actual runtime of an algorithm for specific problems of interest is crucially determined by prefactors \cite{elfving2020will},
as well as by other technical considerations, 
such as the amount of input and output data that need to be pre-computed and post-processed respectively, and moved between the classical and quantum computer \cite{von2020quantum}, 
and the precise determination of the accuracy of quantum algorithms by tight numerical bounds as opposed to loose inequalities \cite{childs2019nearly}.
Furthermore, quantum walks and LCU based algorithms require additional quantum resources, especially ancillae and controlled operations, 
which are challenging aspects when implementation on near-term quantum devices is considered.

In our view, such observations pinpoint the need of detailed comparative studies, conducted on a diverse range of chemical problems,
with exhaustive cross-checks, validations and tests between classical and quantum algorithms. Such detailed comparisons can 
precisely establish by numerical studies the regime (i.e. the number of spin-orbitals $M$, simulation time $t$, and target accuracy $\varepsilon$)
where quantum walks and LCU based algorithms become less expensive than product formulas, and of course of other algorithms for classical computers \cite{elfving2020will},
and provide rigorous benchmarks for assessing the current state of the art in quantum simulation, for measuring its progress, 
and for developing new and improved techniques.
We now describe some applications of algorithms for Hamiltonian dynamics simulation, that can represent occasions for such comparative studies.

{Another important research direction aims at combining elements of each family of algorithms and fine-tuning their implementation for specific problems. Relevant examples are a procedure \cite{berry2015hamiltonian} to use LCU on the steps of the quantum walk described in Sec.~\ref{sec:qw} and the observation \cite{berry2018improved,poulin2018quantum} that, for the purpose of estimating energies, 
the steps of the quantum walk Eq.~\eqref{eq:qw} are sufficient, rather than the Hamiltonian evolution unitary constructed from the quantum walk.}

\paragraph{Time-dependent observables and correlation functions.}
A natural application of the algorithms outlined in Section~\ref{sec:dynamics} is the calculation of time-dependent electrostatic properties, presented in Section~\ref{sec:electron_dynamics}.
The monitoring and control of electronic motion in atoms and molecules in real time has been made possible by advances in laser technology \cite{hentschel2001attosecond,kienberger2002steering,bucksbaum2003ultrafast,fohlisch2005direct},
and can be addressed on classical computers by a variety of numerical methods \cite{klamroth2003laser,saalfrank2005laser,krause2005time,krause2007molecular,nest2005multiconfiguration,daley2004time,schollwock2011density,xie2019time,dahlen2007solving}.
Achieving this goal requires solving the time-dependent Schr\"{o}dinger equation with $\hat{H}(t) = \hat{H}_0 + \hat{{\bf{d}}} \cdot {\bf{E}}(t)$,
where $\hat{{\bf{d}}}$ is the dipole operator and ${\bf{E}}(t)$ is a time-dependent electric field,
and computing electronic densities and polarizabilities over the time-evolved wavefunction.

\begin{figure}[h!]
\includegraphics[width=0.35\textwidth]{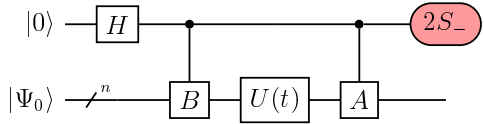}
\caption{Quantum circuit to measure the correlation function $\langle \Psi_0 | \hat{U}^\dagger(t) \hat{A} \hat{U}(t) \hat{B} | \Psi_0 \rangle$ 
between unitaries $\hat{A}$, $\hat{B}$, with $S_- = \frac{X+iY}{2}$.} 
\label{fig:green}
\end{figure}

Another important set of observables based on the simulation of Hamiltonian dynamics are time-dependent correlation functions, such as dipole-dipole correlation functions
$d_{\mu\nu}(t) = \langle \Psi_0 | \hat{d}_\mu(t) \hat{d}_\nu | \Psi_0 \rangle$. The electronic structure Hamiltonian
and the dipole operators are mapped onto linear combinations of Pauli operators, {$\hat{d}_\mu = \sum_j c_{j\mu} \hat{P}_j$}, reducing the computation of dipole-dipole correlation functions
to that of Pauli operators. The quantum circuit \cite{ortiz2001quantum,somma2002simulating,somma2003quantum} 
for the calculation of time-dependent correlation functions between unitary operators $\hat{A}$, $\hat{B}$ is shown in Fig.~\ref{fig:green}.
The calculation of time-dependent correlation functions is also an important occasion to study, demonstrate, benchmark and improve the performance of 
quantum circuits comprising ancillae and controlled operations, an important and recurring theme in quantum simulation \cite{chiesa_2019,francis_2020,sun2021quantum,cohn2021quantum}.

\paragraph{Quantum phase estimation (QPE).}

QPE is one of the most important subroutines in quantum computation. 
It serves as a central building block for many quantum algorithms, 
and implements a measurement for essentially any Hermitian operator $\hat{H}$,
most notably the Hamiltonian, through an algorithmic implementation
of von Neumann's general measurement scheme \cite{neumann1955mathematical}.
Basic quantum-mechanical measurements are performed by decomposing 
$\hat{H}$ into easily measurable terms, measuring each term separately, 
and collating results. 
QPE, on the other hand, prepares an eigenstate of the Hermitian operator 
to be measured in one register, and stores the corresponding eigenvalue 
in a second register.
As such, QPE only requires a single shot and has the zero-variance property
(if $\Psi$ is an eigenfunction of $\hat{H}$ with eigenvalue $E_0$, then
the QPE measurement of $\hat{H}$ returns $E_0$ with probability 1).

More technically, QPE is used to estimate the eigenvalue $u = e^{i 2\pi \theta}$, 
$0 < \theta < 1$, corresponding to the eigenvector $|u \rangle$ of a unitary $\hat{U}$ \cite{cleve1998quantum,kitaev1995quantum}. 
In the context of quantum simulation \cite{aspuru2005simulated,lanyon2010towards,o2016scalable,o2019quantum,cruz2020optimizing} 
$\hat{U}$ is a controllably accurate approximation of {$e^{-i \lambda \hat{H}}$ and $\lambda$ a suitable rescaling factor}, 
and QPE is thus used to compute eigenvalues corresponding to ground and excited state of $\hat{H}$.
A simple but important observation is that eigenvalues of $\hat{H}$ do not lie between $0$ and $2\pi$, 
however one can scale and shift the Hamiltonian to an operator $\hat{H}^\prime = 2 \pi (\hat{H}-E_1)/(E_2-E_1)$, 
where $E_1$/$E_2$ is a lower/upper bound for the lowest/highest eigenvalue of $\hat{H}$,
satisfying such a condition.

\begin{figure}[h!]
\includegraphics[width=0.5\textwidth]{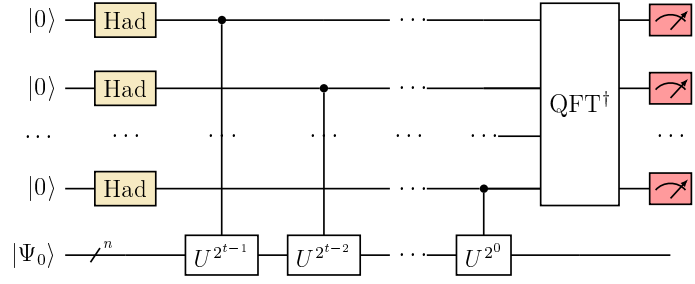}
\caption{Quantum circuit for the quantum phase estimation algorithm.}
\label{fig:qpe}
\end{figure}

There are two main strategies for algorithmic QPE: 
the first makes use of the gate expensive inverse quantum Fourier transform (QFT) and, in an ideal quantum computer, could work with a single measurement,
the second uses shallower circuits \cite{kitaev1995quantum,griffiths1996semiclassical,dobvsivcek2007arbitrary} but requires multiple measurements and classical post-processing.
The former implementation of the QPE algorithm is described by the quantum circuit in Figure \ref{fig:qpe}.

A register of qubits prepared in $\ket{u}$ is coupled to $t$ ancillae prepared in $|0 \rangle^{\otimes t}$.
Hadamard gates and controlled powers of $\hat{U}$ are applied, 
and a subsequent inverse quantum Fourier transform \cite{nielsen2002quantum,benenti2004principles}
leads to the final state 
\begin{equation}
| \Psi \rangle 
= 
\sum_{z=0}^{2^t-1} \left( \sum_{k=0}^{2^t-1} \frac{ e^{ \frac{2 \pi i z (2^t \theta-k)}{2^t} } }{2^t } \right) \ket{z} \otimes \ket{u}
= 
\sum_{z=0}^{2^t-1} \Delta(z,\theta) \, \ket{z} \otimes \ket{u}
\quad.
\end{equation}
If $2^t \theta = k_0$ for some integer $k_0 = 0 \dots 2^t-1$, then $| \Psi \rangle = \ket{k_0} \otimes \ket{\psi}$, 
and measuring the ancillae yields the binary representation of $k_0 = 2^t \theta$ with probability $1$.
Otherwise, the probability distribution $p(z) = |\Delta(z,\theta)|^2$ is concentrated around the integer $k_0$ closest to $2^t \theta$,
$p(k_0) \geq 4/\pi^2 \simeq 0.4$ \cite{cleve1998quantum}.
$p(k_0)$ can be increased to $1-\varepsilon$ increasing the number of ancillae to $t = \mathcal{O}(\log \varepsilon^{-1})$ \cite{cleve1998quantum}.

A large body of work is dedicated to the optimization of QPE,
from reducing the number of measurements needed to estimate eigenvalues in implementations not based on QFT \cite{svore2013faster},
to methodologies for simultaneously determining multiple eigenvalues based on a classical time-series analysis \cite{o2019quantum,somma2019quantum},
to optimizing the implementation of QPE on contemporary quantum devices \cite{cruz2020optimizing,mohammadbagherpoor2019improved}.

\paragraph{Adiabatic state preparation (ASP).} This technique approximates the ground state of an interacting system.
The Hamiltonian is written as $\hat{H} = \hat{H}_0 + \hat{H}_1$, where the eigenvalues and eigenvectors of $\hat{H}_0$ can be easily determined and encoded on a classical or a quantum computer
(in chemistry, a natural choice is $\hat{H}_0 = \hat{F} + \langle \Psi_{\mathrm{HF}} | \hat{H} - \hat{F} | \Psi_{\mathrm{HF}} \rangle$, where $\hat{F}$ is the Fock operator and $\Psi_{\mathrm{HF}}$ the Hartree-Fock state)
and a curve of operators $\hat{H}(s)$, $0 \leq s \leq 1$, with $\hat{H}(s=0) = \hat{H}_0$ and $\hat{H}(s=1) = \hat{H}$, for example the segment $\hat{H}(s) = \hat{H}_0 + s \hat{H}_1$.
The adiabatic theorem \cite{born1928beweis,kato1950adiabatic,messiah1962quantum,avron1999adiabatic,teufel2003adiabatic,jordan2008quantum} 
states that, under opportune conditions, the solution of the Schr\"{o}dinger equation
\begin{equation}
\label{eq:asp}
i \frac{d}{dt} \ket{\Psi_t} = \hat{H}(t/T) \ket{\Psi_t} 
\quad,\quad
0 \leq t \leq T
\quad,\quad
\ket{\Psi_0} = \ket{\Phi_0(0)}
\quad,
\end{equation}
where $\ket{\Phi_0}$ is the ground state of $\hat{H}_0$, converges to the ground state $\ket{\Phi_0(1)}$ of $\hat{H}$ in the large $T$ limit,
\begin{equation}
\lim_{T \to \infty} | \Psi(T) \rangle = | \Phi_0(1) \rangle
\quad.
\end{equation}
APS uses an operation suited for the quantum computer, the simulation of time evolution, to approximate Hamiltonian ground states.
The method was originally proposed to address combinatorial optimization problems \cite{farhi2000quantum,farhi2001quantum} 
and later generalized to chemistry problems \cite{du2010nmr,babbush2014adiabatic,veis2014adiabatic},
as well as to a model of quantum computation, equivalent to the circuit model \cite{kempe2006complexity,nagaj2007new,aharonov2008adiabatic} 
and featuring interesting robustness properties against coherent and incoherent errors \cite{childs2001robustness}.
An important point is the following: the adiabatic theorem states that the time $T$ to approximate the ground state with accuracy $\varepsilon$ scales as 
$\varepsilon^{-1} F(\gamma(s),f_1(s),f_2(s))$, where $F$ is a functional of the spectral gap $\gamma(s)$ of $\hat{H}(s)$,
and of the norms $f_k(s) = \| \frac{d^k \hat{H}}{ds^k} \|(s)$ of the first and second derivative of $\hat{H}$ along the adiabatic path \cite{jordan2008quantum}.
The simulation time, and thus the computational cost of ASP, are especially connected with $\gamma(s)$:
if such a quantity remains constant, or decreases as $1/\mbox{poly}(M)$ where $M$ is the number of spin-orbitals of the system, 
ASP is polynomially expensive \cite{van2001powerful,jansen2007bounds}.
Otherwise, it can be exponentially expensive, in accordance with the QMA nature of the ground state problem.

\subsection{Simulation of Hamiltonian eigenstates}

The problem of computing Hamiltonian eigenpairs, $\hat{H} \ket{\Phi_\mu} = E_\mu \ket{\Phi_\mu}$ has enormous importance in chemistry 
(see e.g. the applications in Sections \ref{sec:es} and \ref{sec:applications_of_time_evolution}).
This problem lies in the QMA complexity class, and thus the existence of quantum algorithms outperforming their classical counterparts is not expected.
However, heuristic quantum algorithms can, for certain structured problems, produce accurate approximations of ground and selected excited states at polynomial cost.
In this Section, we review heuristic quantum algorithms for the computation of approximate Hamiltonian eigenstates and eigenvalues.

\subsubsection{Variational quantum algorithms}

Variational quantum algorithms (VQAs) have recently emerged as a widely used strategy to approximate Hamiltonian eigenstates/eigenvalues on quantum computers \cite{cao2019quantum,cerezo2020variational,bauer2020quantum,bharti2021noisy}, in part due to the fact that VQAs can be designed to operate within the limitations of
contemporary quantum hardware.
To define and implement a VQA, one first considers a parametrized wavefunction (or Ansatz) such as 
\begin{equation}
\label{eq:vqa_0}
\ket{\Psi(\theta)} = \hat{U}(\theta) \ket{\Psi_0}
\quad,\quad
\hat{U}(\theta) = \hat{u}_{n_g-1}(\theta_{n_g-1}) \dots \hat{u}_{0}(\theta_{0})
\quad,
\end{equation}
where $\ket{\Psi_0}$ is an initial wavefunction and the $\{ \hat{u}_k \}_k$ are parameterized unitaries.
VQAs typically operate preparing the parametrized Ansatz \eqref{eq:vqa_0} on a quantum computer, executing a circuit, measuring the obtained state,
and updating the parameters $\theta$ according to a classical optimization algorithm, based on the results of such measurements.
The rather abstract structure of VQAs materializes in a wealth of particular implementations,
which can be roughly classified in two families: variational quantum optimization (VQO) and variational quantum simulation (VQS) algorithms.
The former approximate target states by minimizing a suitable cost function, and the latter approximate dynamical processes
corresponding to curves in a Hilbert space by minimizing a suitable action functional.

\subsubsection{The variational quantum eigensolver}

\begin{figure}[h!]
\centering
\includegraphics[width=0.62\textwidth]{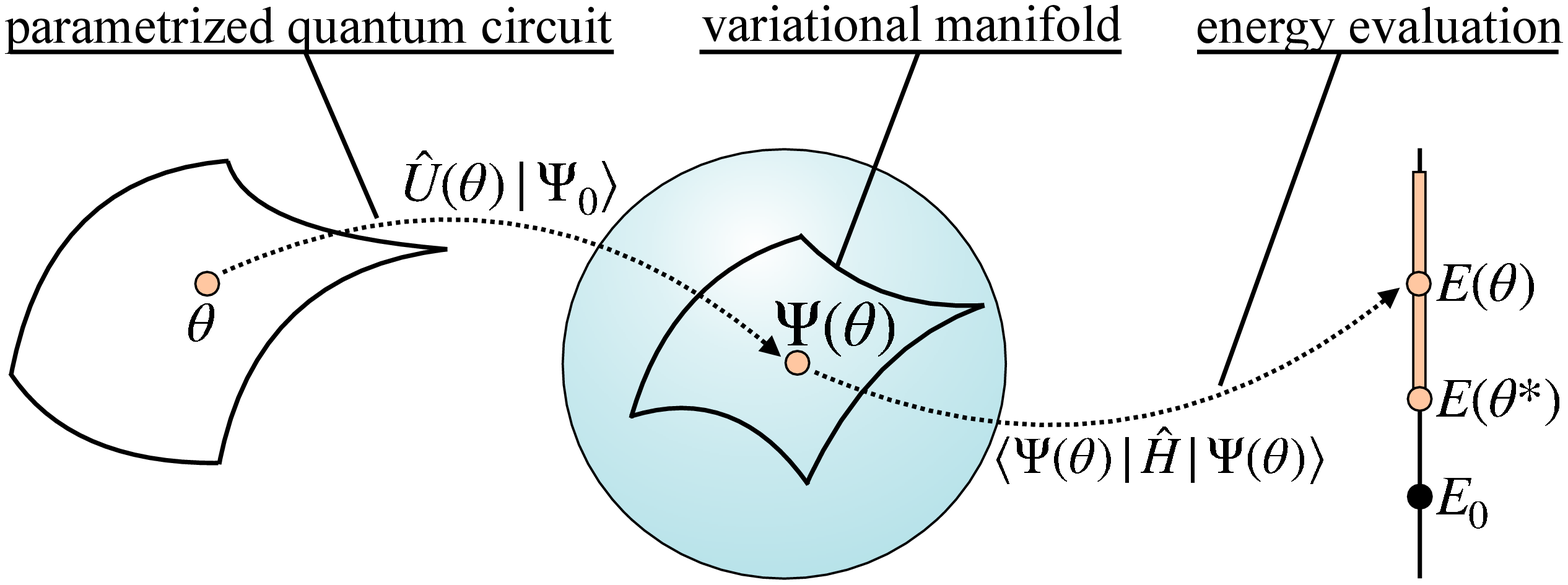}
\caption{Schematics of the VQE algorithm. 
A parametrized wavefunction is produced applying a parametrized unitary to an initial wavefunction.
The set of wavefunctions accessible to the VQE algorithm is a manifold in the qubit Hilbert space,
and the wavefunction minimizing the energy is chosen as an approximation of the ground state wavefunction.
}
\label{fig:vqe}
\end{figure}

A prominent example of a VQO is the variational quantum eigensolver (VQE) \cite{farhi2014quantum,peruzzo2014variational,mcclean2016theory,romero2018strategies}, schematized in Fig.~\ref{fig:vqe},
which approximates the ground-state energy and wavefunction  of a Hamiltonian $\hat{H}$ by minimizing the energy
$E(\theta) = \langle \Psi(\theta) | \hat{H} | \Psi(\theta) \rangle$ which, according to the variational principle, is an upper bound for the ground-state energy of $\hat{H}$.
In the context of VQE, the quantum computer is used to efficiently evaluate $E(\theta)$ and, in some implementations, its first and second derivatives \cite{mcclean2016theory,parrish2019hybrid,schuld2019evaluating,mitarai2020theory,kottmann2021feasible}.
A simple and important aspect of VQO methods in general and VQE in particular, is that evaluating the energy on a quantum computer yields a statistical estimate, $E(\theta) \sim \mu \pm \sigma$,
and thus optimizers used to update parameters have to take into account the statistical nature of quantum measurements \cite{guerreschi2017practical}.
Example of such optimizers are the Simultaneous Perturbation Stochastic Approximation (SPSA) \cite{spall2005introduction,hirokami2006parameter,s2013stochastic},
ADAptive Moment estimation (ADAM) \cite{kingma2014adam},
and the Quantum Natural Gradient (QNG) \cite{stokes2020quantum}.
SPSA is a stochastic optimization method where parameters are updated as $\theta_{n+1} = \theta_n - a_n \, g(\theta_n)$,
where $g(\theta_n)$ is an estimate of the gradient of the cost function obtained from random perturbation vectors of length $c_n = c \, n^{-\gamma}$,
and the step length is defined as $a_n = a \, n^{-1}$. Careful optimization of the hyperparameters $a,c,\gamma$ is key to an efficient optimization \cite{kandala2017hardware}.
ADAM is a first-order gradient-based optimization of stochastic objective functions, based on adaptive estimates of lower-order moments, characterized by a simple implementation, modest memory requirements,
and the ability to dynamically select a step size by maintaining a history of past gradients.
Within QNG, the optimization dynamics is interpreted as moving in the steepest descent direction with respect to the quantum information geometry, corresponding to the real part of the Fubini-Study metric tensor.

The accuracy and computational cost of VQE calculations are determined by the underlying Ansatz: on the one hand it should contain an accurate approximation to the ground state,
on the other hand one desires circuits that can be easily executed on a quantum computer,
with a well-behaved and rapidly convergent optimization of variational parameters \cite{mcclean2018barren,akshay2020reachability}.

Two such Ans\"{a}tze  are reviewed in the following paragraphs.

\begin{figure}[t!]
\includegraphics[width=0.7\textwidth]{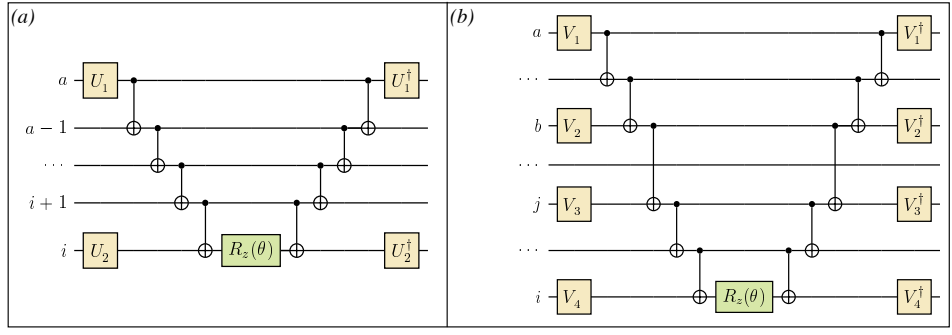}
\caption{Circuits for the exponentiation of the single (a) and double 
(b) excitation operators in Eq.~\eqref{eq:cc_basic}. The repeated 
units across several qubits are shown in dashed lines, and the Clifford
unitaries are {$(U_1,U_2)=\{(A_Y,A_X)$, $(A_X,A_Y)\}$ and 
$(V_1,V_2,V_3,V_4) = (A_X,A_X,A_Y,A_X)$, $(A_Y,A_X,A_Y,A_Y)$,
$(A_X,A_Y,A_Y,A_Y)$, $(A_X,A_X,A_X,A_Y)$, $(A_Y,A_X,A_X,A_X)$,
$(A_X,A_Y,A_X,A_X)$, $(A_Y,A_Y,A_Y,A_X)$, $(A_Y,A_Y,A_X,A_Y) \}$}
respectively.}
\label{fig:q-uccsd}
\end{figure}

\paragraph{Quantum unitary coupled-cluster (q-UCC).} The ground state is approximated \cite{bartlett1989alternative,peruzzo2014variational,Moll_2018,romero2018strategies,albash2018adiabatic,o2016scalable} 
by an exponential Ansatz,
\begin{equation}
\label{eq:q_uccsd}
| \Psi \rangle = e^{ \hat{T} - \hat{T}^\dagger } | \Psi_0 \rangle
\quad,\quad
\hat{T} = \sum_{l=1}^{k} \hat{T}_l
\quad,\quad
\hat{T}_l = \sum_{a_0 \dots a_{l-1}} \sum_{i_0 \dots i_{l-1}} t^{a_0 \dots a_{l-1}}_{i_0 \dots i_{l-1}} \; \crt{a_0} \dots \crt{a_{l-1}} \dst{i_{l-1}} \dots \dst{i_0} 
\quad,
\end{equation}
where $| \Psi_0 \rangle$ is a reference Slater determinant and $\hat{T}$ is a linear combination of up to $k$ particle-hole excitation operators that promote
electrons from the hole (occupied) to the particle (unoccupied) orbitals of the reference state $| \Psi_0 \rangle$.
In many studies, $\hat{T}$ is limited to single and double particle-hole excitations, defining the so-called q-UCCSD Ansatz.
Standard coupled cluster theory is naturally implemented on a classical device, while its unitary variant is naturally implemented on a quantum device.
Eq.~\eqref{eq:q_uccsd} is factored into a product of exponentials of Pauli operators, using product formulas presented in Section \ref{sec:dynamics},
\begin{equation}
e^{ \hat{T} - \hat{T}^\dagger } \simeq \prod_\mu e^{ \theta_\mu \left( \hat{t}_\mu - \hat{t}_\mu ^\dagger \right) }
\quad.
\end{equation}
Such a factorization is not unique, and optimal parameterization of fermionic wavefunctions via q-UCC has been recently explored \cite{evangelista2019exact}. For ease of implementation, Eq.~\eqref{eq:q_uccsd} can be approximated \cite{barkoutsos2018quantum} by a product of exponentials of individual
one- and two-body operators,
\begin{equation}
\label{eq:cc_basic}
e^{ \hat{T} - \hat{T}^\dagger } \simeq \prod_{ia} 
e^{ \theta^a_i ( \crt{a} \dst{i} - \crt{i} \dst{a} ) }
\prod_{ijab} e^{\theta^{ab}_{ij} ( \crt{a} \crt{b} \dst{j} \dst{i} - \crt{i} \crt{j} \dst{b} \dst{a} ) }
\quad.
\end{equation}
In the JW representation, the exponentials in Eq.~\eqref{eq:cc_basic}
are represented by the quantum circuits in Figure \ref{fig:q-uccsd},
each of which require $\mathcal{O}(M)$ $\mathsf{CNOT}$ gates and have
depth  $\mathcal{O}(M)$.
While the resulting quantum circuit has polynomial depth and number of
gates, the $\mathcal{O}(M^5)$ scaling of a basic implementation is 
sufficiently deep to limit implementations of q-UCCSD on today's 
quantum hardware. 

Note that, when using a VQE solver, all Ans\"{a}tze will give an energy 
that is above the true energy of the system provided that all the 
appropriate properties/symmetries (e.g. number of particles, $S^2$, 
$S_z$) are maintained.

\begin{figure}[h!]
\includegraphics[width=0.55\textwidth]{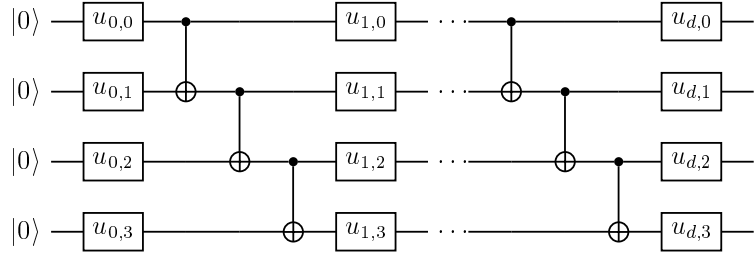}
\caption{Example of a hardware-efficient Ansatz for a 4-qubit problem, defined by layers of single-qubit gates $u_{i,j}$ interspersed with $\CNOT$ gates acting 
on adjacent qubits on a device with linear connectivity.}
\label{fig:hardware_efficient}
\end{figure}

\paragraph{Hardware efficient Ans\"{a}tze.} Hardware efficient Ans\"{a}tze are typically designed to be experimentally feasible, 
because they are based on the realizable demands on connectivity and gate operations of a given chip.
An example is shown in Fig.~\ref{fig:hardware_efficient}, and consists of alternating layers of arbitrary single-qubit gates and
an entangling gate.
While it is not guaranteed that such Ans\"{a}tze contain good approximations to the state of interest, they enable important and 
conceptually insightful simulations on contemporary quantum devices.

{It is worth emphasizing that some heuristic methods, epitomized by q-UCC, are based on hierarchies of increasingly more general (and therefore more accurate and expensive) Ans\"{a}tze. As such, they offer the possibility to systematically converge results towards exact quantities, provided enough computational resources are available.}

The design of compact Ans\"atze that join hardware efficiency and chemical insight is an active and valuable research area.
An important progress in this regard was the development of {schemes like qubit coupled-cluster \cite{ryabinkin2018qubit} and ADAPT-VQE \cite{Grimsley2019,tang2021qubit}. In ADAPT-VQE,}
a pool of operators $\{ \hat{A}_i \}_i$ is chosen in advance,
the ground state is approximated by the parametrized wavefunction $| \Psi_\theta \rangle = \exp( \theta_n \hat{A}_{i_n} ) \dots \exp( \theta_1 \hat{A}_{i_1} ) | \Psi_0 \rangle$,
where the angles are optimized variationally,
and pool operators are appended to the circuit based on the value taken by the energy gradient $g_i = \langle \Psi_\theta | [\hat{H} , \hat{A}_i] | \Psi_\theta \rangle$.
Numerical simulations showed ADAPT-VQE can improve over q-UCCSD in terms of the accuracy achievable for a given circuit depth.

\subsubsection{Variational quantum simulation}
\label{sec:vqs}

Unlike VQO algorithms, which approximate a specific point of the Hilbert space by minimizing a cost function, 
VQS algorithms aim at approximating a curve in the Hilbert space of a system (corresponding to a dynamical process) by a curve of time-dependent parametrized wavefunctions, $\ket{ \Psi_{\theta_t} }$.
The flow of such a wavefunction is mapped to the evolution of the parameters $\theta_t$, which takes the form of a differential equation \cite{mcardle2019variational,yuan2019theory}.
For example, to variationally simulate Hamiltonian dynamics, parameters are evolved to make the vector
\begin{equation}
\ket{\Delta} = \frac{d}{dt} \ket{ \Psi_{\theta_t} } + i \hat{H} \ket{ \Psi_{\theta_t} }
=
\sum_k \frac{d \theta_k}{dt} \ket{ \Psi^k_{\theta_t} } + i \hat{H} \ket{ \Psi_{\theta_t} }
\quad,\quad
\ket{ \Psi^k_{\theta_t} } = \frac{\partial}{\partial {\theta_k}} \ket{ \Psi_{\theta_t} }
\quad,
\end{equation}
vanish. McLachlan's variational principle, i.e. minimization of $\| \Delta \|^2$, leads to the differential equation
\begin{equation}
\label{eq:working_eq_vqs}
b_r = \sum_k A_{rk} \frac{d\theta_k}{dt}
\quad,\quad
A_{rk} = \mbox{Re}\left( \langle \Psi^r_{\theta_t} | \Psi^k_{\theta_t} \rangle \right) 
\quad,\quad
b_r = \mbox{Im} \left( \langle \Psi^r_{\theta_t} | \hat{H} | \Psi_{\theta_t} \rangle \right)
\quad,
\end{equation}
defining $d \theta/dt$.
The quantities $A$, $b$ are measured on the quantum computer \cite{mcardle2019variational,yuan2019theory}, 
and the classical computer uses such information to compute $d \theta/dt$ and to update the parameters $\theta$.
VQS algorithms are especially useful to carry out simulations of Hamiltonian dynamics on contemporary quantum hardware:
while the Schr\"odinger equation can be solved by converting $\exp(-it\hat{H})$ into a quantum circuit, as discussed in Section \ref{sec:dynamics},
the depth of such a circuit generally increases polynomially with simulation time $t$.
Like their VQO counterparts, VQS algorithms have heuristic nature, as they assume that the quantum state is represented by an Ansatz quantum circuit with fixed depth and structure at any time $t$.
Depending on the problem of interest and the structure of the Ansatz, VQS algorithms may thus give inaccurate results.

\subsubsection{Quantum diagonalization algorithms}

\begin{figure}[h!]
\includegraphics[width=0.3\textwidth]{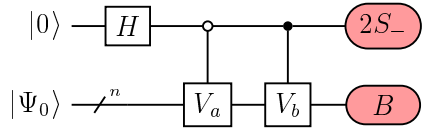}
\caption{Quantum circuit for the measurement of $B_{ab} = \langle {\bf{v}}_a | \hat{B} | {\bf{v}}_b \rangle$.}
\label{fig:lanczos}
\end{figure}

In classical electronic structure, the search for ground and excited Hamiltonian eigenstates can be tackled relying on diagonalization methods.
Given a Hamiltonian acting on a Hilbert space $\mathcal{H}$ with dimension $N$, diagonalization methods are based on the synthesis of a 
collection of vectors $\{ | \vett{v}_a \rangle \}_{a=0}^{d-1}$ that form a basis for a $d$-dimensional subspace of $\mathcal{H}$.
Once these vectors are available, the overlap and Hamiltonian matrices 
\begin{equation}
\label{eq:diag_eigen}
S_{ab} = \langle \vett{v}_a | \vett{v}_b \rangle 
\quad,\quad
H_{ab} = \langle \vett{v}_a | \hat{H} | \vett{v}_b \rangle
\quad,
\end{equation}
are computed, and the eigenvalue problem $H \vett{c}_\mu = E_\mu S \vett{c}_\mu$ is solved, 
to determine approximate eigenvalues $E_\mu$ and eigenvectors $| \psi_\mu \rangle = \sum_a c_{a\mu} \, | \vett{v}_a \rangle$ of $\hat{H}$ \cite{lanczos1950iteration,davidsorq1975theiterative,morgan1986generalizations}.
Recently, a number of quantum diagonalization algorithms have been conceived \cite{mcclean2017hybrid,colless2018computation,huggins2020non,motta2020determining,ollitrault2020quantum,parrish2019quantum,huggins2020non,stair2020multireference,jamet2021krylov}, 
that generate vectors $| \vett{v}_a \rangle$ applying suitable operators to an initial state $| \Psi_0 \rangle$,
\begin{equation}
| \vett{v}_a \rangle = \hat{V}_a | \Psi_0 \rangle \quad,\quad a = 0 \dots d-1 \quad .
\end{equation}
When the $\hat{V}_a$ are unitary operators, the matrix elements Eq.~\eqref{eq:diag_eigen} can be computed with the so-called Hadamard test circuit \cite{somma2002simulating,aharonov2009polynomial}, shown in Fig.~\ref{fig:lanczos}.
An alternative approach is to compile the operators $\hat{V}_a^\dagger \hat{V}_b$ and $\hat{V}_a^\dagger \hat{H} \hat{V}_b$ into linear combinations of Pauli operators, 
and measure them with the techniques described in Section \ref{sec:quantum_computers}, provided the number of Pauli operators grows polynomially with qubit number.
Depending on the number and the nature of the vectors $| \vett{v}_a \rangle$, and on the structure of the problem at hand, 
quantum diagonalization algorithms can offer a polynomially expensive route to accurate approximation for ground and selected excited states.
Examples of quantum diagonalization algorithms are briefly described in the remainder of this Section.

\paragraph{Quantum subspace expansion (QSE).} In this algorithm, excited states are defined \cite{mcclean2017hybrid,colless2018computation,huggins2020non} by the Ansatz 
$| \psi_\mu \rangle = \sum_a c_{a\mu} \hat{E}_a | \Psi_0 \rangle$, where $\{ \hat{E}_a\}_a$ is a set of pre-defined excitation operators.
Common choices are Pauli operators of weight at most $k$ and fermionic operators of rank at most $k$,
\begin{equation}
\begin{split}
P_k 
&= 
\left\{ \hat{\sigma}_{{\bf{m}}} = \hat{\sigma}_{m_{n-1}} \otimes \dots \otimes \hat{\sigma}_{m_0} 
\;,\;
\sigma_{m_i} \neq \mathbbm{1}
\mbox{ for at most $k$ indices }
\right\}
\quad, \\
F_k 
&= 
\left\{ \crt{i_0} \dots \crt{i_{l-1}} \dst{j_{l-1}} \dots \dst{j_0} \;,\; l \leq k \right\} 
\quad.
\end{split}
\end{equation}

\paragraph{Quantum filter diagonalization (QFD).} This technique \cite{parrish2019quantum,huggins2020non,stair2020multireference,cohn2021quantum}
projects the Hamiltonian on a subspace spanned by a set of non-orthogonal quantum states generated via approximate time evolution,
$\ket{{\bf{v}}_k} = \exp(-i k \Delta t \hat{H}) \, \ket{\Psi_0}$.
QFD can be regarded as a quantum computational equivalent of classical filter diagonalization \cite{neuhauser1990bound,neuhauser1994circumventing},
from which it inherits the connection with the Lanczos algorithm. 
Furthermore, it draws a profound connection between the two central problems of quantum simulation, 
namely BQP-complete Hamiltonian simulation and QMA-complete ground-state search.
{Finally, it provides} a compelling framework to apply and test algorithms for approximate and variational time evolution,
because QFD requires time evolution to generate a set of linearly independent states, a goal that can be achieved 
with less conservative approximations.

\paragraph{Quantum equation of motion (q-EOM).}

QSE computes total energies of ground and excited states, which can be subtracted to yield excitation energies.
One route to compute excitation energies directly is the q-EOM method, 
well established in classical simulations and recently extended to quantum computation \cite{Rowe,Ganzhorn,ollitrault2020quantum,gao2020applications,barison2020quantum}.
In this framework, excitation energies are computed as 
\begin{equation}
\Delta E_\mu = \frac{ \langle  \Psi_0 | [\hat{O}_\mu,\hat{H},\hat{O}_\mu^\dag] | \Psi_0 \rangle }{ \langle  \Psi_0 | [ \hat{O}_\mu ,\hat{O}_\mu^\dag] | \Psi_0 \rangle }
\quad,
\end{equation} 
where $2 [\hat{O}_\mu,\hat{H},\hat{O}_\mu^\dag] = [[\hat{O}_\mu,\hat{H}],\hat{O}_\mu^\dag] + [\hat{O}_\mu,[\hat{H},\hat{O}_\mu^\dag]]$
and $\hat{O}_\mu$ is an excitation operator expanded on a suitable basis. 
{Commutators are introduced for several reasons. First, they can be used to compute energy differences directly rather than total energies: in fact, if $\Psi_0$ is an eigenstate of $\hat{H}$ with eigenvalue $E_0$, then $\hat{H} \hat{O}_\mu^\dagger |\Psi_0 \rangle = [ \hat{H} , \hat{O}_\mu^\dagger ] |\Psi_0 \rangle + E_0 \hat{O}_\mu^\dagger |\Psi_0 \rangle$
and thus $\hat{O}_\mu^\dagger |\Psi_0 \rangle$ is an eigenstate of $\hat{H}$ with eigenvalue $E_\mu$ if $[ \hat{H} , \hat{O}_\mu^\dagger ] |\Psi_0 \rangle = (E_\mu - E_0) \hat{O}_\mu^\dagger |\Psi_0 \rangle = \Delta E_\mu \hat{O}_\mu^\dagger |\Psi_0 \rangle$. Second, they can project the Schr\"{o}dinger equation onto a subspace of relevant electronic wavefunctions \cite{Rowe} and finally, the rank (number of electronic excitations) of a commutator is lower than the rank of a product, which has a beneficial impact on the
computational cost.}
The variational problem of finding the stationary points of $\Delta E_\mu$ leads to a generalized eigenvalue equation, the solutions of which are the excited-state energies.

\paragraph{Quantum Lanczos (qLANCZOS).} In this algorithm, 
imaginary-time evolution (ITE)
\begin{equation}
\ket{{\bf{v}}_k} = \frac{ e^{-k \Delta \tau \hat{H}} \ket{\Psi_0} }{\| e^{-k \Delta \tau \hat{H}} \ket{\Psi_0} \|}
\quad,\quad
k = 0 \dots d-1
\quad,
\end{equation}
is used to construct the subspace. {QFD and qLANCZOS are examples of methods projecting the Schr\"{o}dinger equation in a time series basis.}
ITE is a non-linear and non-unitary map, and thus is not naturally simulatable on a digital quantum computer.
Various approaches to achieve this goal have been proposed, ranging from LCU-based to variational  
(see Sections \ref{sec:taylor}, \ref{sec:vqs}). 
Ref.~\cite{motta2020determining} introduced an alternative approach to apply ITE on a quantum computer, termed quantum ITE or QITE, 
which is free from ancillae and controlled operations as well as from high-dimensional parameter optimizations. In the QITE method,
a single step of ITE under a geometrically local term $\hat{h}_m$ of the Hamiltonian is approximated by a unitary,
$e^{- \Delta\tau \hat{h}_m} \ket{\Psi} \propto e^{i\hat{A}_m} \ket{\Psi}$, where $\hat{A}_m = \sum_i x_{im} \hat{P}_{im}$,
the operators $\hat{P}_{im}$ act on a neighborhood of the domain of $\hat{h}_m$, and the coefficients $x_{im}$
are determined from local measurements \cite{motta2020determining}.
While initial estimates in a limited set of problems show QITE to be resource-efficient compared to variational methods
\cite{motta2020determining,sun2021quantum,yeter2020a,gomes2020,yeter2020b,kamakari2021digital},
an extensive numerical understanding of its performance and cost across different problems remains to be developed.

\subsubsection{Applications of variational algorithms, and open problems}

The design and improvement of heuristic algorithms for Hamiltonian eigenpair approximation 
is one of the most active research areas at the interface between quantum chemistry and quantum computation.
Some of the directions of current and future research include: the creation of new Ans\"{a}tze based on chemical notions,
the extension of properties accessible to heuristic algorithms, and the economization of calculations.

The relationship between standard coupled-cluster $e^{\hat{T}} \ket{\Psi_0}$ and unitary coupled-cluster Eq.~\eqref{eq:q_uccsd} 
exemplifies how quantum computation can offer occasions to revive and reinterpret concepts and techniques from quantum chemistry.
In both cases, the use of a cluster expansion is motivated by the qualitative accuracy of mean-field theory,
and quantum computation provides a compelling framework to explore the theoretical and numerical differences between these theories \cite{Cooper2010,harsha2018difference,evangelista2019exact,lee2018generalized}
especially in statically correlated situations.
Circuits inspired by the q-UCCSD hierarchy but which directly substitute the fermionic field operators for spin ladder operators \cite{ryabinkin2018qubit} have been suggested, 
as well as circuits based on the use of specific building blocks \cite{o2019generalized,anselmetti2021local,barison2020quantum,matsuzawa2020jastrow} and symmetry preserving Ans\"{a}tze \cite{gard2020efficient}.

The VQE algorithms has been extended to the calculation of excited states \cite{higgott2019variational,ibe2020calculating}
and the optimization of molecular orbitals \cite{mizukami2020orbital,sokolov2020quantum} by suitable modification of the cost function;
to properties other than the ground-state energy \cite{rice2021quantum,sokolov2021microcanonical} by evaluation of suitable operators on the VQE wavefunction;
and have been integrated in the workflow of solid-state chemistry \cite{choudhary2021quantum,ma2020quantum} , 
transcorrelated Hamiltonian \cite{motta2020quantum,mcardle2020improving} 
and quantum embedding \cite{rubin2016hybrid,dhawan2020dynamical,metcalf2020resource,kawashima2021efficient,rossmannek2021quantum} calculations.
{Modifying the VQE cost function has also been proposed as a technique to improve the quality or the convergence of ground-state simulations \cite{stair2021simulating,kuroiwa2021penalty,ryabinkin2018constrained}.}

On the front of algorithm optimization, considerable effort has been devoted to reducing the measurement cost \cite{gonthier2020identifying}, for example
by simultaneously measuring commuting subsets of the Pauli operators needed for the cost function 
\cite{wecker2015progress,jena2019pauli,izmaylov2019unitary,jena2019pauli,kubler2020adaptive,zhao2020measurement}, {leveraging amplitude amplification \cite{wang2021minimizing}},
as well as adopting machine-learning techniques to extract more information from a given measurement dataset \cite{torlai2020precise,hadfield2020measurements,hillmich2021decision}.
{In the context of variational quantum algorithms for time evolution, 
important research directions are related with the simplification of the working equations \eqref{eq:working_eq_vqs}, and the economization of the quantum measurement required by the simulation. For example, VQS techniques based on minimizing the distance (or maximizing the overlap) between states evolved in time exactly and variationally have been proposed \cite{barison2021efficient,benedetti2021hardware}, along with techniques to economize quantum measurements introducing causal light-cone structure in the Ansatz \cite{foss2021holographic,benedetti2021hardware,kattemolle2021variational}. Generalizing these techniques to chemical Hamiltonians is a compelling research direction, at the interface between quantum algorithms for physics and chemistry.}

{Furthermore, while molecular simulations generally require handling the N-electron wavefunction, recently proposed approaches have instead focused on the expression of the ground-state energy as a functional of the two-electron reduced density matrix \cite{boyn2021quantum,mazziotti2021quantum}. While recent results 
have indicated that they represent a promising direction
for efficient molecular quantum simulations, additional research is needed to assess their full potential.}

\section{Error mitigation techniques for near-term quantum devices}
\label{sec:hardware}

Until recently, executing quantum algorithms was only a theoretical possibility. 
Recent advances have made quantum computing devices available to the scientific community \cite{ibm2020services,rigetti2020services},
and computational packages to design and implement quantum algorithms \cite{smith2016practical,Qiskit,mcclean2020openfermion}.

\begin{figure}[h!]
\includegraphics[width=\textwidth]{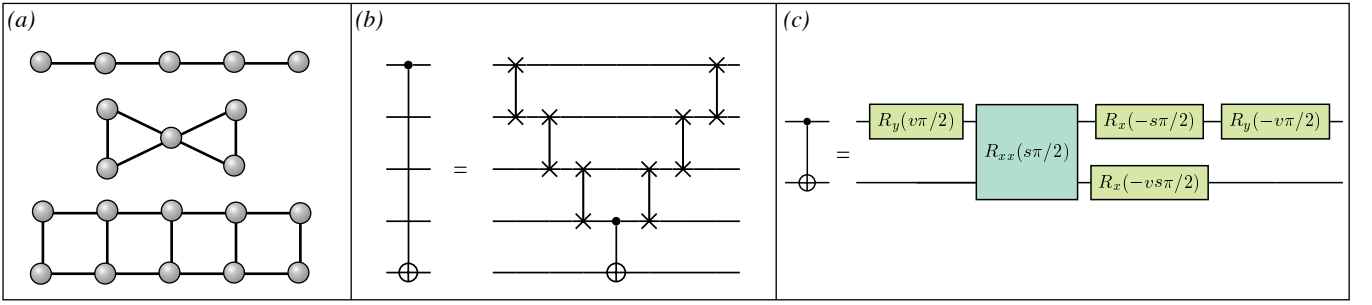}
\caption{Schematics (a) of chips with linear (top), bow-tie (middle) and ladder (bottom) topology. Implementation (b) of a $\CNOT$ gate between two non-adjacent qubits in a device with linear topology results in a significant overhead of $\mathsf{SWAP}$ gates.
Implementation (c) of a $\CNOT$ gate on a device where native gates are single-qubit rotations and $R_{xx}$ gates, {with $s,v = \pm 1$ (adapted from \cite{maslov2017basic})}.
}
\label{fig:hardware}
\end{figure}

\begin{table}[h!]
\begin{tabular}{ccc}
\hline\hline
reference & systems & number of qubits \\
\hline
\cite{peruzzo2014variational}  & \ce{HeH+} & 2 \\
\cite{o2016scalable,kandala2017hardware}                & \ce{H2}      & 2 \\
\cite{kandala2017hardware}    & \ce{BeH2} & 6 \\
\cite{kandala2017hardware,rice2021quantum} & \ce{LiH} & 4 \\
\cite{nam2020ground}             & \ce{H2O} & 4 \\
\cite{gao2021computational}   & \ce{LiO2} dimer & 2 \\
\cite{mccaskey2019quantum} & \ce{NaH}, \ce{RbH}, \ce{KH} & 4 \\
\cite{google2020hartree}         & \ce{H12}   & 12 \\
\cite{eddins2021doubling}       & \ce{H2O} & 5 \\
\hline\hline
\end{tabular}
\caption{Some recent experiments on contemporary quantum hardware, aimed at simulating molecular systems .}
\label{table:recent_simulations}
\end{table}

Based on a broad range of architectures, such as superconducting \cite{krantz2019quantum,devoret2004superconducting} and trapped ion \cite{bruzewicz2019trapped,brown2021materials} qubits,
such devices are capable of carrying out quantum computations of chemical systems on a limited scale, as exemplified in Table \ref{table:recent_simulations}, for a variety of technical reasons.
In particular:
($i$) they comprise less than 100 qubits which, as seen in Section \ref{sec:mapping_to_qubits}, limits the number of electrons and orbitals that can be simulated,
($ii$) not all pairs of qubits are physically connected, so that entangling gates have to be limited to adjacent qubits in the topology of the chip, 
or implemented incurring an overhead of $\mathsf{SWAP}$ gates, see Fig.~\ref{fig:hardware},
($iii$) each device has a set of native gates, dictated by its architecture and manipulation techniques \cite{rigetti2010fully,chow2011simple,yan2018tunable}; 
while such gates are universal (see Section \ref{sec:universality}),
every gate in a quantum circuit has to be compiled into a product of native gates,
($iv$) quantum hardware is subject to decoherence and imperfect implementation of quantum operations.
Errors occurring on a quantum device can be classified into coherent (unitary noise processes) and incoherent (non-unitary noise processes),
\begin{equation}
i \hbar \frac{d \rho_{\textrm{hw}} }{dt} = [ \hat{H}_{\textrm{hw}}(t) + \hat{H}_{\textrm{c}}(t) , \rho_{\textrm{hw}} ] + \mathcal{L}_{ \textrm{i} }( t, \rho_{\textrm{hw}} )
\quad,
\end{equation}
where $\rho_{\textrm{hw}}$ is the density operator of the quantum hardware, 
and $\hat{H}_{\textrm{c}}(t)$ and $\mathcal{L}_{ \textrm{i} }$ generate unitary and non-unitary evolution respectively.
Coherent errors are exemplified by over- or under-rotation in qubit control pulses and qubit cross-talking,
and examples of incoherent errors are the following single-qubit amplitude damping, phase damping and depolarization processes, respectively \cite{nielsen2002quantum}
\begin{equation}
\mathcal{L}_{\textrm{a}}(\rho) = \gamma_{\textrm{a}} \; \mathcal{V}_{S_-}(\rho)
\quad,\quad
\mathcal{L}_{\textrm{p}}(\rho) = \gamma_{\textrm{p}} \; \mathcal{V}_{Z}(\rho)
\quad,\quad
\mathcal{L}_{\textrm{d}}(\rho) = \gamma_{\textrm{d}} \; \sum_m \mathcal{V}_{\sigma_m}(\rho)
\quad,
\end{equation}
where $\mathcal{V}_{A}(\rho) = A \rho A^\dagger - A^\dagger A \rho - \rho A^\dagger A$.
Relaxation and dephasing processes occur on time scales $T_1$ and $T_2$ respectively, called qubit decoherence times.
Deeper quantum circuits comprising more entangling gates have significantly larger biases and statistical uncertainties, 
due to the accumulation of coherent and incoherent errors,
an effect that is especially pronounced when the execution time of the circuit is comparable with decoherence times of the qubits.
Fully addressing the decoherence problem requires an advanced set of techniques, 
namely fault-tolerant quantum computation via quantum error correction \cite{peres1985reversible,shor1995scheme,steane1996error,nielsen2002quantum,fowler2012surface},
which in turn requires extremely low error rates for qubit operations as well as a significant overhead of physical qubits \cite{fowler2012surface,preskill2018quantum}.
Enhancing the capabilities of near-term quantum computing hardware thus requires techniques to mitigate errors 
without requiring any additional quantum resources \cite{kandala2018extending,maciejewski2020mitigation,bharti2021noisy}.

The emergent nature of quantum devices requires quantum chemists and quantum information scientists to conduct synergistic research,
so that algorithmic implementations understand and leverage the nature, limitations and features of quantum hardware,
and the benchmarking and development of quantum hardware is driven by chemical applications.
Conducting quantum simulations of chemical systems and designing algorithms for contemporary quantum hardware 
is also an important occasion to understand, characterize and mitigate errors induced by experimental imperfections.
In the remainder of this section, we will describe some {emerging} techniques for mitigation of readout and gate errors.

\paragraph{Readout error mitigation.}
In theoretical considerations about quantum information protocols, a quantum device is often assumed to perform unbiased measurements.
In practice, this assumption is often violated due to experimental imperfections and decoherence. Since measurements are a central part
of any quantum simulation, this observation led to the development of readout error mitigation techniques.

Let us denote by ${\bf{p}}_{\textrm{ideal}}$ the exact probability distribution for the outcomes of a quantum measurement,
and ${\bf{p}}_{\textrm{exp}}$ the probability distribution actually measured on the quantum hardware.
The relationship between ${\bf{p}}_{\textrm{ideal}}$ and ${\bf{p}}_{\textrm{exp}}$ is in general captured by a complicated multidimensional function.
However, if noise affecting measurements is weak, an accurate approximation can be obtained assuming the 
relation between the two probability distributions is given by a linear map,
${\bf{p}}_{\textrm{exp}} = \Lambda \, {\bf{p}}_{\textrm{ideal}} + {\bf{\Delta}}$.
The elements of $\Lambda$ and ${\bf{\Delta}}$ can be estimated by 
choosing a set of calibration circuits $\{ c_k \}_k$ such that ${\bf{p}}_{\textrm{ideal}}(k)$ can be computed analytically (e.g. circuits comprising a single layer of $X$ gates), 
measuring the observable of interest over these circuits,
obtaining probability distributions ${\bf{p}}_{\textrm{exp}}(k)$, 
and minimizing the distance $\Lambda_0,{\bf{\Delta}}_0 = \mbox{argmin}_{\Lambda,{\bf{\Delta}}} \sum_k \| {\bf{p}}_{\textrm{exp}}(k) - \Lambda {\bf{p}}_{\textrm{ideal}}(k) - {\bf{\Delta}} \|$. 
The observable of interest is measured over a different state outside the calibration set, 
and the ideal probability distribution is reconstructed from the experimental one as 
${\bf{p}}_{\textrm{ideal}} \simeq \Lambda^{-1}_0 \, \left[ {\bf{p}}_{\textrm{exp}} - {\bf{\Delta}}_0 \right]$. 
This method is solely based on classical post-processing and, although its cost scales exponentially with qubit number, 
suitable Ans\"{a}tze on the structure of the pair $\Lambda_0,{\bf{\Delta}}_0$ can still give accurate maps at polynomial cost.
Numerical studies have analyzed the impact of finite statistics (at the stage of estimation of probability distributions) on the protocol,
and confirmed its approximate validity on a variety of publicly available prototypes of quantum chips \cite{temme2017error,kandala2018extending,maciejewski2020mitigation}. 

\begin{figure}[h!]
\includegraphics[width=0.65\textwidth]{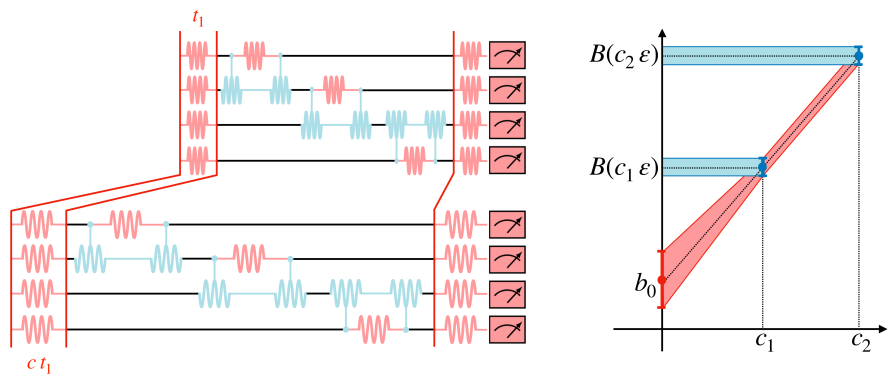}
\caption{Left: a measurement of the expectation value after rescaled state preparation is equivalent to a measurement under an amplified noise strength, 
if the noise is time-translation invariant. Right: illustration of gate error mitigation based on a first-order Richardson extrapolation to the zero-noise limit. This 
highlights that the variance of the mitigated estimate $b_0$ is dependent on the variance of the unmitigated measurements, and the stretch factors $c_i$.}
\label{fig:gate_error_mitigation}
\end{figure}

\paragraph{Gate error mitigation.}
Recent work \cite{temme2017error,li2017efficient,endo2018practical} has shown that the accuracy of computation based off expectation values of quantum observables, such as variational quantum algorithms, 
can be enhanced through an extrapolation of results from a collection of varying noisy experiments.
Any quantum circuit can be expressed in terms of evolution under a time-dependent drive Hamiltonian $\hat{H}_{\textrm{hw}}(t) = \sum_\alpha J_\alpha(t) \hat{P}_\alpha$ acting on the quantum hardware,
where $\hat{P}_\alpha$ represents some Hermitian operator of the quantum hardware and $J_\alpha(t)$ the strength of the associated interaction. 
The expectation value $B(\varepsilon)$ of an observable of interest over the state prepared by the drive $\hat{H}_{\textrm{hw}}(t)$ in the presence of noise can be expressed as a power series around its zero-noise value,
\begin{equation}
B(\varepsilon) = b_0 + \sum_{k=1}^n b_k \, \varepsilon^k + \mathcal{O}(\varepsilon^{n+1})
\quad,
\end{equation}
where $\varepsilon$ is a small noise parameter, and  the coefficients in the expansion $b_k$ are dependent on specific details of the noise model.
The primary objective of gate error mitigation techniques is to experimentally obtain improved estimates to $b_0$ despite using noisy quantum hardware. 
Assuming noise is time-translationally invariant a possible strategy \cite{kandala2018extending}, sketched in Fig.~\ref{fig:gate_error_mitigation}, 
is to perform a collection of experiments with stretched pulses,
$J_\alpha(t) \to c_i^{-1} J_\alpha(c^{-1}_i t)$ corresponding to noise strengths $c_i \varepsilon$, computing the corresponding expectation values
$B_i(\varepsilon) = B(c_i \varepsilon) = b_0 + \sum_k b_k (c_i \varepsilon)^k$, and extracting $b_0$ using a Richardson extrapolation \cite{richardson1911approximate}.
This protocol, demonstrated for a variety of applications within and beyond molecular electronic structure \cite{kandala2017hardware},
proved able to enhance the computational capabilities of quantum processors based on superconducting architectures, with no additional quantum resources or hardware modifications,
which makes it very compelling for practical implementations on near-term hardware.
It is important to notice that implementing the Richardson extrapolation protocol requires a profound understanding and control of the gates used in the circuit,
which in turn motivated less general but more easily implementable schemes \cite{dumitrescu2018cloud,rice2021quantum}.
Furthermore, unlike quantum error correction, gate error mitigation techniques do not allow for an indefinite extension of the computation time, 
and only provide corrections to expectation values, without correcting for the full quantum mechanical probability distributions.

{There exist other gate error mitigation techniques, that achieve cancellation of errors, for example, 
by introducing quantum gates implementing unitary transformations generated by the symmetries of the system \cite{tran2021faster}, or by resampling randomized circuits according to a quasi-probability distribution \cite{temme2017error}.}

\begin{figure}[h!]
\includegraphics[width=0.28\textwidth]{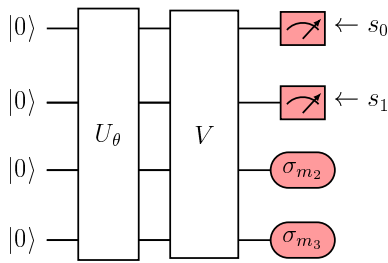}
\caption{
Schematics of a post-selection procedure. The transformation $\hat{V}$ achieves simultaneous measurement of the stabilizer generators, transformed to single-qubit operators
acting on the first $k=2$ qubits, from which the stabilizer parities $s_0,s_1$ are read. 
The other qubits are measured in $X$, $Y$ or $Z$ basis depending on the Pauli string of interest, and measurement outcomes with the undesired parity are discarded.
}
\label{fig:post_selection}
\end{figure}

\paragraph{Post-selection.}
To mitigate the effect of hardware noise on the measurement results, one can also process hardware data by error-mitigation methods such as post-selection.
When a Hamiltonian has $\mathbbm{Z}_2$ symmetries, as discussed in Sec.~\ref{sec:fermions_second}, a wavefunction encoded on a quantum computer can be
written as {$| \Psi \rangle = \hat{U}^\dagger \sum_{{\bf{s}}} c_{ {\bf{s}} } | \Phi_{\bf{s}} \rangle \otimes | {\bf{s}} \rangle$}, where the stabilizer parities ${\bf{s}}$ label irreps of the symmetry group.
In absence of noise, wavefunctions should have an intended stabilizer parity ${\bf{s}}_0$, to avoid mixing different irreps.
However, during execution of the circuit, gate errors and qubit decoherence can induce nonzero overlap of the qubit state with subspaces of undesired parity. 
Post-selection can mitigate these undesirable effects by discarding measurement outcomes with the wrong parity \cite{bonet2018low,mcardle2019error},
as exemplified in Fig.~\ref{fig:post_selection}.
Compared against qubit reduction, post-selection requires more qubits, but typically retains the compact and efficient representation of fermionic and other operators as qubit operators,
which the transformation $\hat{U}$ typically undoes.
In the context of molecular electronic structure, post-selection is particularly appealing when simulations are performed in the fermionic Fock space. 
While Fock space representations are often elected for ease of implementation, electronic structure wavefunctions have well-defined particle number and spin.
When the JW representation is used in conjunction with low-rank decompositions of the Hamiltonian \cite{huggins2021efficient,cohn2021quantum}, 
several such constants of motion quantities can be efficiently measured simultaneously with the one- and two-body contributions to the Hamiltonian,
thereby projecting the electronic wavefunction into an eigenspace of constants of motion labeled by desired eigenvalues.

\section{Conclusion and Outlook}
\label{sec:conclusions}

In this work, we explored {emerging} quantum computing algorithms for chemistry.  
In our discussion, we emphasized that quantum computers are special purpose machines, capable of solving certain structured problems with a polynomial
amount of resources.
These problems, exemplified by the simulation of Hamiltonian dynamics,
can be projected to benefit from quantum algorithms.
For other problems, exemplified by the simulation of Hamiltonian eigenpairs, 
quantum algorithms are based on heuristic approximation schemes.

We reviewed quantum algorithms for the simulation of Hamiltonian dynamics 
(product formulas, quantum walks, and LCU-based algorithms) and of 
Hamiltonian eigenstates (variational and diagonalization algorithms), 
highlighting some applications and open problems.

Given the emergent nature of quantum computation, a number of properties 
of quantum algorithms need to be characterized by systematic numerical 
studies over a set of diverse chemical problems, especially accuracy and computational cost.
This characterization is needed for both heuristic 
(to help understand, establish and refine the underlying approximations)
and non-heuristic algorithms 
(in order to determine when and how to apply them).

Chemists can contribute to this effort in many different ways.
First, they can help design sets of chemical systems that interpolate
between small, simple (e.g. H$_2$) and large, realistic cases (e.g. enzymes).
Achieving this goal can help demonstrate algorithms on today's devices,
as well as lead to a more systematic understanding of their scalability and accuracy.
Chemists can contribute also to the continuous development of new heuristics,
by collaborating with quantum information scientists to optimally represent
chemical wavefunctions and observables in terms of quantum circuits and 
measurements respectively.

\section*{Acknowledgment}

We thank {T. J. Lee, D. Maslov, H. Nakamura, A. Mezzacapo, and D. W. Berry} 
for helpful feedback on the manuscript.

\appendix

\section{Glossary}

\begin{table}[h!]
\centering
\resizebox{\textwidth}{!}{
\begin{tabular}{|ll|ll|}
\hline\hline
acronym & meaning & acronym & meaning \\
\hline
2QR & two-qubit reduction & QITE & quantum imaginary-time evolution (ITE) \\
ADAM & adaptive moment estimation & qLANCZOS & quantum Lanczos  \\
ASP & adiabatic state preparation & QMA & quantum Merlin-Arthur \\
BK & Bravyi-Kitaev & QM/MM & quantum mechanics / molecular mechanics approach \\
BQP & bounded-error quantum polynomial time & QNG & quantum natural gradient  \\
CCSD & coupled-cluster (CC) with singles and doubles & QPE & quantum phase estimation  \\
CCSD(T) & CCSD with perturbative estimate to connected triples & QSE & quantum subspace expansion  \\
CI & configuration interaction & QSP & quantum signal processing \\
DFT & density functional theory & q-UCC & quantum unitary CC  \\
FCI & full CI & q-UCCSD & q-UCC with singles and doubles \\
HF & Hartree-Fock & STO-6G & minimal basis where 6 primitive Gaussian orbitals \\
JW & Jordan-Wigner & & are fit to a Slater-type orbital (STO) \\ 
LCU & linear combination of unitaries & SPSA & simultaneous perturbation stochastic approximation  \\
OAA & oblivious amplitude amplification & VQA & variational quantum (VQ) algorithm  \\
q-EOM & quantum equation of motion & VQE & VQ eigensolver  \\
QFD & quantum filter diagonalization & VQO & VQ optimization  \\
QFT & quantum Fourier transform & VQS & VQ simulation  \\

\hline\hline
\end{tabular}
}
\caption{Glossary of acronyms used throughout the present work.}
\label{tab:glossary}
\end{table}


%

\end{document}